\documentclass[%
reprint,
superscriptaddress,
w%amsmath,
amssymb,amsmath,
]
{revtex4-1}
\usepackage[english]{babel} %francais, polish, spanish, ...
\usepackage [table]{xcolor}
\usepackage{graphicx, upgreek, subfigure} %%For loading graphic files
\usepackage{float}
\usepackage{textcomp}
\usepackage{graphicx}% Include figure files
\usepackage{dcolumn}% Align table columns on decimal point
\usepackage{bm}% bold math
\usepackage[mathlines]{lineno}% Enable numbering of text and display math
\usepackage{scalerel,amssymb}
\usepackage{physics}
\usepackage{array}

\newcommand{\sitwoeight}{$^{28}$Si}
\newcommand{\sitwonine}{$^{29}$Si}
\newcommand{\ttwo}{$T_2$}

\newcommand{\sz}{$S_{\rm z}$}
\newcommand{\sx}{$S_{\rm x}$}
\newcommand{\dfdb}{$\partial f/\partial B_0$}
\newcommand{\natsi}{$^{\rm nat}$Si}

\newcommand{\SI}{Supplementary Materials}
\begin{document}

\begin{abstract}
%{\color{red}Tellurium impurities in silicon are deep double donors with attractive properties for use as silicon based spin qubits or in spin ensemble quantum memories. When singly ionised, typically achieved via co-doping with B, and cooled to cryogenic temperatures, the Te$^+$ system consists of an electron spin bound to the spin-1/2 Te nucleus with a large hyperfine coupling of 3.496~GHz. 
%A zero-field magnetic clock transition occurs at this frequency ---such transitions are insensitive to magnetic noise and exhibit enhanced electron spin coherence times.

%We study the spins of group-VI $^{125}$Te$^+$ donors implanted into natural Si at depths of 20 and 300~nm, examining spin activation yield, relaxation ($T_1$) and coherence times ($T_2$). We show how the presence of a 3.5~GHz `clock transition' at zero magnetic field can be used to extend spin coherence times to over 1~ms, as well as narrow the inhomogeneous spin linewidth half width half maximum to 0.6~MHz. Shallow group-V donors, such as phosphorus, are typically ionised very close to the surface due to band-bending caused by surface states, or metal interfaces in surface nanoelectronic devices, removing their electron spin. We show how such ionisation caused by band-bending can be used to generate Te$^+$ (with electron spin $S=1/2$), and that coherence times of these near-surface donors are comparable to those ionised in the bulk using boron compensation. Overall, these results show that $^{125}$Te$^+$ has many features that will be of interest for silicon-based spin qubits and ensemble quantum memories.

Impurity spins in crystal matrices are promising components in quantum technologies, particularly if they can maintain their spin properties when close to surfaces and material interfaces. Here, we investigate an attractive candidate for microwave-domain applications, the spins of group-VI $^{125}$Te$^+$ donors implanted into natural Si at depths as shallow as 20~nm. We show that surface band-bending can be used to ionise such near-surface Te to spin-active Te$^+$ state, and that optical illumination can be used further to control the Te donor charge state. We examine spin activation yield, spin linewidth, relaxation ($T_1$) and coherence times (\ttwo) and show how a zero-field 3.5~GHz `clock transition' extends spin coherence times to over 1~ms, which is about an order of magnitude longer than other near-surface spin systems.

\end{abstract}

% \title
% {$^{125}$Tellurium deep-donors in silicon - a platform for nanoscale quantum devices}
% \title
% {Spin coherence of near-surface
% $^{125}$Te$^+$ deep-donors in silicon}
\title
{Near-surface $^{125}$Te$^+$ spins with millisecond coherence lifetime}
%{Spin coherence of near-surface ionised $^{125}$Te$^+$ donors in silicon}

\author{Mantas~\v{S}im\.{e}nas}
\altaffiliation{These authors have contributed equally to this work} 
\affiliation{London Centre for Nanotechnology, UCL, 17-19 Gordon Street, London, WC1H 0AH, UK}

\author{James~O'Sullivan}
\altaffiliation{These authors have contributed equally to this work} 
\affiliation{London Centre for Nanotechnology, UCL, 17-19 Gordon Street, London, WC1H 0AH, UK}

\author{Oscar~W.~Kennedy}
\altaffiliation{These authors have contributed equally to this work} 
\affiliation{London Centre for Nanotechnology, UCL, 17-19 Gordon Street, London, WC1H 0AH, UK}

\author{Sen Lin}
\affiliation{Department of Physics, Centre for Quantum coherence and The Hong Kong Institute of Quantum Information Science and Technology, The Chinese University of Hong Kong, Hong Kong, China}

\author{Sarah Fearn}
\affiliation{Department of Materials, Imperial College London, London SW7 2BX, UK}

\author{Christoph~W.~Zollitsch}
\affiliation{London Centre for Nanotechnology, UCL, 17-19 Gordon Street, London, WC1H 0AH, UK}

\author{Gavin Dold}
\affiliation{London Centre for Nanotechnology, UCL, 17-19 Gordon Street, London, WC1H 0AH, UK}

\author{Tobias Schmitt}
\affiliation{Institute for Semiconductor Nanoelectronics, Peter Grünberg Institute 9, Forschungszentrum Jülich and RWTH Aachen University, Germany}

\author{Peter Sch\"uffelgen}
\affiliation{Institute for Semiconductor Nanoelectronics, Peter Grünberg Institute 9, Forschungszentrum Jülich and RWTH Aachen University, Germany}

\author{Ren-Bao Liu}
\affiliation{Department of Physics, Centre for Quantum coherence and The Hong Kong Institute of Quantum Information Science and Technology, The Chinese University of Hong Kong, Hong Kong, China}

\author{John~J.~L.~Morton}
\altaffiliation{jjl.morton@ucl.ac.uk} 
\affiliation{London Centre for Nanotechnology, UCL, 17-19 Gordon Street, London, WC1H 0AH, UK}
\affiliation{Department of Electrical and Electronic Engineering, UCL, Malet Place, London, WC1E 7JE, UK}

\maketitle
Donor spins in nanoscale silicon devices have been shown to be a promising building block for various solid state quantum devices, including atomic qubits~\cite{pla2012single} and quantum memories with coherence times approaching seconds ~\cite{ranjan2020multimode}. Such devices typically contain band discontinuities at silicon/metal and silicon/vacuum interfaces which build electric fields into devices, impacting the charge and spin state of nearby donors. Placing donors close to these interfaces is often important, for example, to increase spin-resonator coupling~\cite{ranjan2020electron}, or couple donor spins to electrostatically-tunable quantum dots ~\cite{urdampilleta2015charge, tosi2017silicon}. Without control of the surface potential (for example, through a metallic top-gate), this typically limits the 
minimum donor-interface distance to tens of nanometres for shallow group-V donor electron spins~\cite{pla2018strain}, whereas deeper donors could be placed closer to electrodes. There is also evidence that, for deeper donors, the electron spin coherence is less strongly influenced by naturally abundant \sitwonine\ spins (as seen by comparing results from P and Bi donors~\cite{witzel2010electron, George2010}).

Singly ionised group-VI chalcogens (S$^+$, Se$^+$, Te$^+$) possess an electron spin $S=1/2$, like the group-V donors, but have much larger ionisation energies~\cite{Grimmeiss1981,nardo2015spin, ludwig1965paramagnetic}. These donors have attracted recent interest due to their optical transitions~\cite{deabreu2019characterization}. 
Through continuous-wave electron spin resonance (ESR) studies~\cite{Grimmeiss1981,niklas1983endor}, $^{125}$Te$^+$ in silicon is known to have a large isotropic hyperfine coupling of $\sim 3.5$~GHz to the  $^{125}$Te nuclear spin ($I=1/2$). 
Singly ionised double donors such as $^{125}$Te$^+$ therefore offer a potential route to maintaining a donor electron spin close to a silicon surface or interface, combined with the presence of a microwave clock transition~\cite{wolfowicz2013atomic} at zero magnetic field.

Another critical challenge encountered when placing electron spins very close to the surface is spin decoherence caused by fluctuating surface defects \cite{ranjan2021spatially,Bluvstein2019}. For example, the coherence time of negatively charged nitrogen-vacancy (NV) centres in diamond drops from a few hundred to tens of microseconds when the surface is approached significantly limiting sensitivity of NVs for nanoscale spin detection and imaging \cite{Wang2016,Fukuda2018,Bluvstein2019}. This stimulates search for other near-surface electron spin centres that are less sensitive to surface-induced decoherence.

Here, we present pulsed ESR measurements of $^{125}$Te$^+$ implanted at depths of 20 and 300~nm in natural silicon. We investigate two different methods to singly ionize $^{125}$Te --- first by co-doping with boron for deep-implanted donors, and second by directly exploiting the band bending arising from Fermi level pinning (FLP) for near-surface donors. We demonstrate superior coherence times of more than 1~ms for near-surface spins as the zero-field clock transition is approached. %Both ionisation methods yielded similar spin coherence times.
We also show how infrared illumination of the shallow-implanted sample improves the ionisation fraction, surpassing that achieved by co-doping. %Overall, our results demonstrate $^{125}$Te to be an attractive candidate for achieving strong spin-resonator coupling with donor spins.

\begin{figure}
    \centering
    \includegraphics[width=\linewidth]{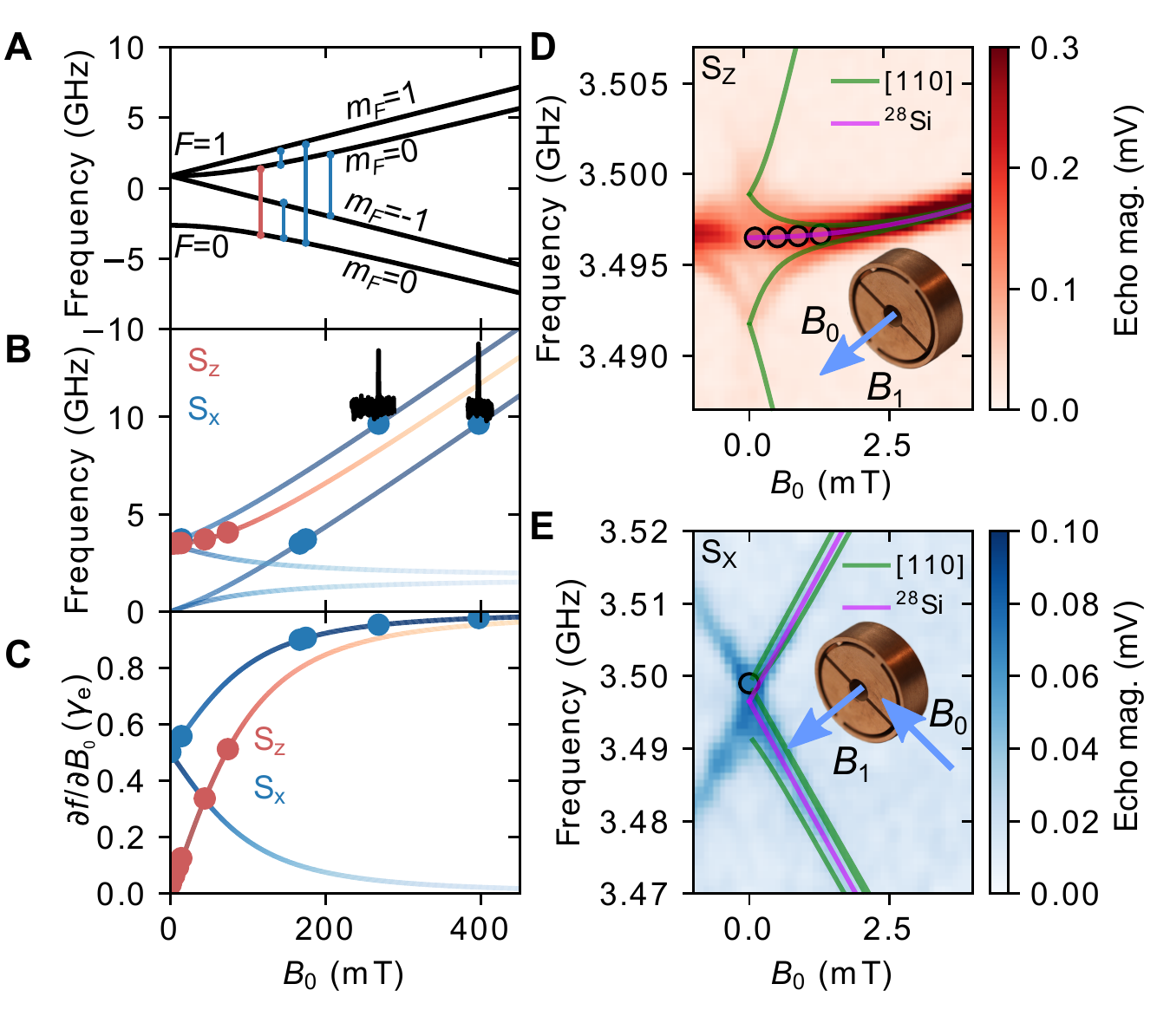}
    \caption{(\textbf{A}) Spin eigenenergies of $^{125}$Te$^+$:Si with \sx~and \sz~transitions marked in blue and red, respectively. 
    %Transitions are labelled by their quantum number $F=S\pm I$ and its projection $m_F$. 
    %Inset are schematics of energy levels and associated microwave transitions at low and high fields. 
    (\textbf{B}) Frequencies of allowed spin transitions, where the line intensity is proportional to the transition probability. Echo-detected spectra obtained at 9.65~GHz are shown in black.  (\textbf{C}) First order magnetic field dependence of the transition frequencies. (\textbf{D},\textbf{E}) Echo-detected spectra of $^{125}$Te$^+$:Si close to zero magnetic field with the cavity oriented to measure (\textbf{D}) \sz~and (\textbf{E}) \sx~transitions. 
    %A one-pixel Gaussian blur is applied. 
    Solid lines show simulations assuming only a hyperfine constant of 3.4965~GHz to $^{125}$Te nucleus, or, in addition, a SHF interaction with a neighbouring [110]-shell \sitwonine\ nuclear spin. All measurements are taken at 10~K. Circles in (\textbf{B})-(\textbf{E}) show fields and transitions, where \ttwo~measurements are performed.}
    \label{fig:Transitions}
\end{figure}

%\section*{Results}
%\subsection*{The Tellurium donor spin system}
The singly ionised $^{125}$Te donor has a single bound electron ($S=1/2$) coupled to the $^{125}$Te nuclear spin ($I=1/2$) via an isotropic hyperfine coupling $A$ (see \SI\ for spin Hamiltonian details).
%
%and is described by the following spin Hamiltonian in  frequency units:
%\begin{equation}
%\label{eq:ESR}
%\mathbf{H}= A\mathbf{S\cdot I}+\gamma_{\rm e}\mathbf{B_0\cdot S}+\gamma_{\rm Te}\mathbf{B_0\cdot I}+\mathbf{H}_{\mathrm{Si}}. 
%\end{equation}
%Here, $\mathbf{B_0}$ is the applied magnetic field, while $\gamma_{\rm e}/2\pi = 28.025(1)$~GHz/T, $\gamma_{\rm Te}/2\pi = -13.5$~MHz/T  are, respectively, the gyromagnetic ratios of the bound electron and $^{125}$Te nucleus. The last term above describes the superhyperfine (SHF) interactions between the electron spin and neighbouring \sitwonine\ nuclei (natural abundance $\sim 4.7$\%):
%\begin{equation}
%\label{eq:shf}
%\mathbf{H}_{\mathrm{Si}}= \sum_i{(\mathbf{S}\cdot \mathbf{A}_{\mathrm{ Si},i}+\gamma_{\rm Si}\mathbf{B_0}) \cdot \mathbf{I}_{\mathrm{Si},i}}.  
%\end{equation}
%Here, $\mathbf{A}_{\mathrm{ Si},i}$ is the SHF coupling to the $i$-th \sitwonine\ nuclear spin, $\mathbf{I}_{\rm Si}$, with gyromagnetic ratio $\gamma_{\rm Si}/2\pi = -8.5$~MHz/T.
%
Fig.~\ref{fig:Transitions}A,B shows the calculated spin transition frequencies ($f$) and probabilities of $^{125}$Te$^+$:Si as a function of magnetic field, neglecting the SHF term.
Transitions are labelled \sz~($\Delta m_F=0$)~or \sx~($\Delta m_F=\pm1$), which are, respectively, driven when the microwave magnetic field component is applied parallel or perpendicular to the static magnetic field. Here, $m_F$ denotes the projection of the quantum number $F=S\pm I$. The ESR experiments described below were performed using a copper loop-gap resonator with adjusted orientation depending on the type of the transition, as shown in insets to Fig.~\ref{fig:Transitions}D,E (see \SI\ for details). 

The first derivative of the transition frequency with respect to the applied field, \dfdb, is an important parameter in determining spin coherence lifetimes and inhomogenous broadening. So-called clock transitions, where \dfdb$~=0$, possess extended coherence times~\cite{Bollinger1985,wolfowicz2013atomic, ortu2018simultaneous, morse2018zero,Vion2002} and narrow linewidths~\cite{wolfowicz2013atomic,Vion2002}. Like most coupled spin systems, $^{125}$Te$^+$ exhibits an \sz~clock transition at zero field (see Fig.~\ref{fig:Transitions}C). Due to the large hyperfine coupling, the clock transition of $^{125}$Te$^+$ occurs in the microwave domain at 3.4965~GHz. In Fig.~\ref{fig:Transitions}D,E we present echo-detected field sweeps (EDFS) of deep-implanted sample for \sz~and \sx~transitions, respectively, whilst varying the frequency of the microwave drive about 3.5~GHz. We resolve the main spin transition as well as the SHF levels revealing perfect agreement with the calculated transition frequency.

\begin{figure}
    \centering
    \includegraphics[width=\linewidth]{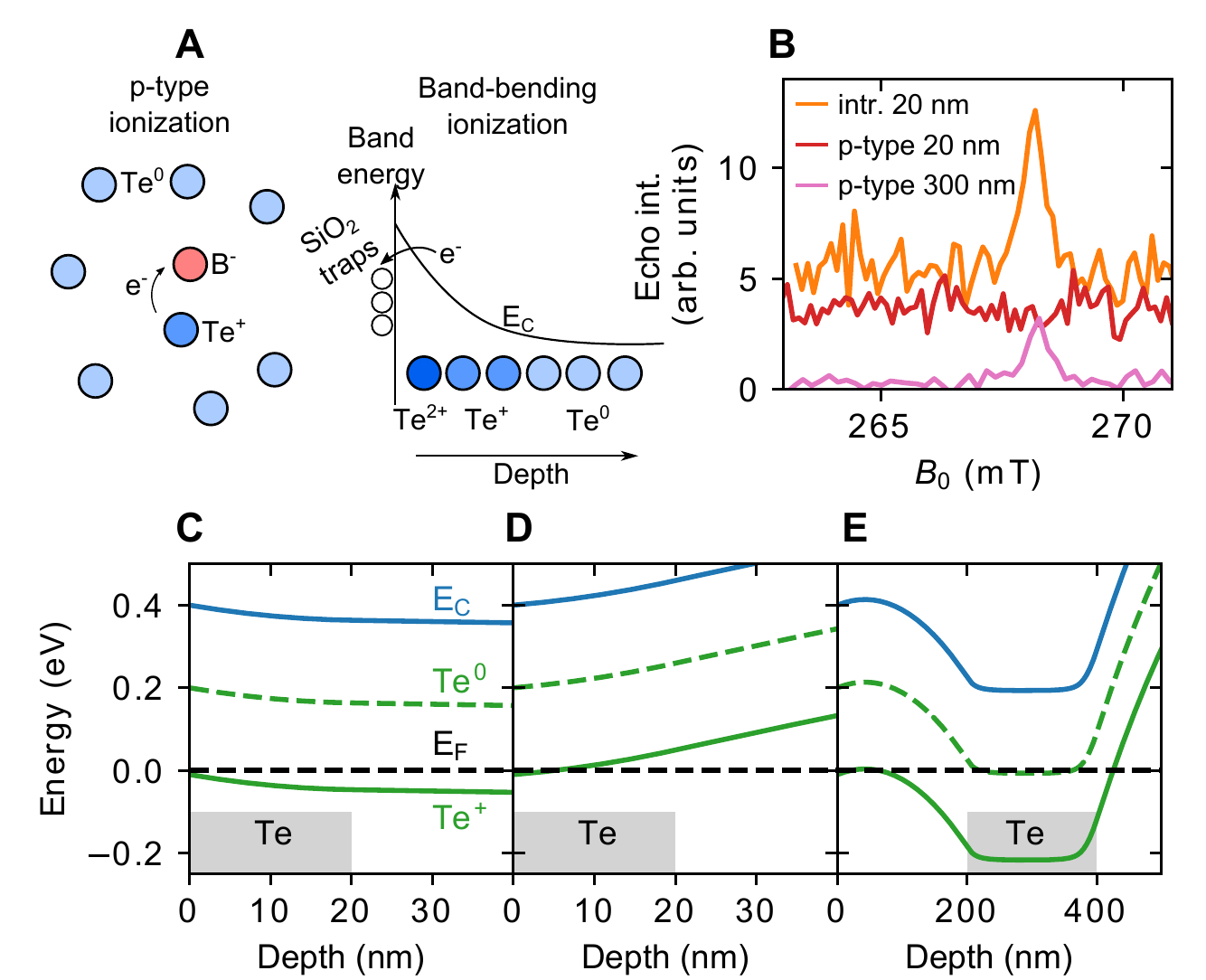}
    \caption{(\textbf{A}) Cartoons showing the different ionisation mechanisms of Te in silicon. (\textbf{B}) EDFSs performed at X-band for \sx~transitions. 
    %The $^{125}$Te$^+$:Si line is present in the 20~keV intrinsic sample and the 800~keV p-type sample, but it is absent in the 20~keV p-type sample. 
    (\textbf{C}-\textbf{E}) Simulated band profiles for the three samples used in this study assuming FLP of 0.4~eV below the conduction band. The simulation results are consistent with EDFSs in (\textbf{B}). (\textbf{C}) Intrinsic wafer with 20~keV implant energy gives all Te singly ionized by surface band bending. (\textbf{D}) p-type wafer with 20~keV implant energy has approximately all Te doubly ionized. (\textbf{E}) p-type wafer with 800~keV implant energy has the Fermi level close to the singly ionized Te level as 1/10th of Te atoms are ionized by sacrificing an electron to the boron acceptors present at 1/10th the density of Te.}
    \label{fig:Ionisation}
\end{figure}

%\subsection*{Activation of tellurium spins}
To be spin active, chalcogens must be incorporated into the silicon lattice (as with group-V donors) and also be singly ionised~\cite{nardo2015spin, ludwig1965paramagnetic}. Using the deep-implanted p-type sample, we first investigated three annealing schedules (5 minutes at 600, 800 and 1000$^\circ$C in dry nitrogen) to incorporate $^{125}$Te into the lattice after implantation. In these samples, ionisation is achieved by co-doping $^{125}$Te ion implanted at 800~keV to a depth of 300~nm with a peak concentration of 10$^{17}$~cm$^{-3}$ (profile shown in \SI) with boron at a concentration of 2.4(2)$\times10^{16}$~cm$^{-3}$ (Fig.~\ref{fig:Ionisation}A). Using an excess of Te, we aim to ionise the majority of the boron into an electron spin-less B$^-$ state to reduce the impact on the Te$^+$ donor spin coherence through spectral diffusion. This necessarily reduces the fraction of Te$^+$ compared to the overall quantity of implanted Te. 

Spin echo measurements at 9.65~GHz show two resonances at the expected magnetic field positions for the \sx\ transitions shown in Fig.~\ref{fig:Transitions}B. By comparing these echo amplitudes to a reference P:Si sample (see \SI\ for details), we determined the activation yield of Te$^{+}$. We observed that activation yield increases from $\sim15$\% to 22\% with annealing temperature increasing from 600$^\circ$C to 1000$^\circ$C (see Table~S1 in \SI). Given that the ionisation mechanism is the same between the samples (B co-doping), we attribute this change to a higher Te incorporation fraction. 
We also observed that the electron spin coherence time \ttwo~obtained at X-band increases with annealing temperature (Table~S1) likely due to healing of spin-active implantation damage at higher annealing temperature. %Secondary ion mass spectrometry (SIMS) depth profiles shown in the \SI\ reveal that annealing at 1000$^\circ$C causes $^{125}$Te to diffuse by $\sim$35~nm in contrast to lower temperature annealing, which causes little diffusion. 
%Further optimization of annealing schedules (e.g.\ longer anneals) would likely allow higher incorporation fraction, healing of implantation damage and control over diffusion. SIMS profiles (Supplementary Materials) show some Te in the surface SiO$_2$, which is not spin active~\cite{Schenkel2009}, accounting for some reduction in the calculated activation. 
%We fixed the annealing schedule at 5~mins 1000$^\circ$C to maximise $T_2$ and incorporation and compared %techniques to singly ionise the $^{125}$Te.

We also implanted intrinsic silicon at a depth of 20~nm using 20~keV implantation energy (SIMS profiles in the \SI) and used surface band bending at the silicon/vacuum interface caused by FLP to ionise Te close to the surface as shown schematically in Fig.~\ref{fig:Ionisation}A. 
Band bending is a ubiquitous effect and also occurs due to Schottky barriers at silicon/metal interfaces and can be controlled by surface treatments~\cite{gleason2013measurement}. % For group-V donors band bending is a concern and may result in a surface depletion layer, where donors are ionised (spinless), which would limit the maximum achievable single spin coupling. In a recent paper, device modelling implied there was no depletion zone around an aluminium electrode and understanding potential profiles around electrodes will be important for optimizing devices~\cite{ranjan2021spatially}.
The EDFS traces in Fig.~\ref{fig:Ionisation}B confirm the generation of Te$^+$ in the shallow-implanted intrinsic sample (as well as the deep-implanted p-type silicon), but there is no signal from the shallow-implanted p-type sample. Note that the number of spins of the shallow-implanted intrinsic sample is much lower compared to the deep-implanted samples (Table S1), resulting in a much weaker ESR signal.%.implying there are few spin-active Te donors in this sample. 

%with FLP 0.4~V below the conduction band consistent with previous measurements for Si/SiO$_2$ interfaces~\cite{dev2003mechanism}

In order to understand the ESR signal strength in the different samples, we self-consistently solved the Schr\"odinger-Poisson equation in one dimension~\cite{tan1990self} and simulated different implants and substrates with various FLP levels.
FLPs in the range of 0.4--0.5~eV give ionisation profiles consistent with the echo amplitudes in Fig.~\ref{fig:Ionisation}B and are in line with literature values for FLP at the silicon/silicon oxide interface~\cite{dev2003mechanism}. 
We show simulations of the band profiles with FLP at 0.4~eV below the conduction band in Fig.~\ref{fig:Ionisation}C-E. Fig.~\ref{fig:Ionisation}C shows simulated band profiles of 10$^{17}$cm$^{-3}$ Te extending 20~nm into intrinsic silicon, where all Te is singly ionised. However, for the shallow-implanted p-type material (Fig.~\ref{fig:Ionisation}D) the simulation predicts predominantly ESR-silent $^{125}$Te$^{2+}$, consistent with the lack of observed spin echo from this sample (Fig. \ref{fig:Ionisation}B). Finally, in the simulation of deep-implanted Te into p-type Si (Fig.~\ref{fig:Ionisation}E) the Fermi level is close to the Te$^+$ level, resulting in Te ionisation commensurate with the boron codoping concentration. See \SI\ for extended simulations.
%We present both of these profiles as a function of FLP level in the Supplementary Materials and show that the shallow p-type sample is singly ionised for 0.2~V~$\lesssim$~FLP $~\lesssim$~0.32~V, and that the ionisation fraction of the deeply implanted sample is determined by the codoping ratio, independent of FLP.
%Increasing the FLP above $0.4$~V produces ESR-silent $^{125}$Te$^{2+}$ at the surface. 

%We also measured the coherence times of the donors in the shallowly implanted intrinsic silicon sample for high \dfdb~\sx~transitions as well as \sz~transitions close to the clock transition (see Fig.~\ref{fig:Coherence}). These coherence times are the same (within experimental errors) as the codoped samples. Despite being near-surface donors ionised by the surface band bending, these spins exhibit similar coherence promising for near-surface or near-interface spins in long-coherence nanoelectronic devices.

\begin{figure}
    \centering
    \includegraphics[width=\linewidth]{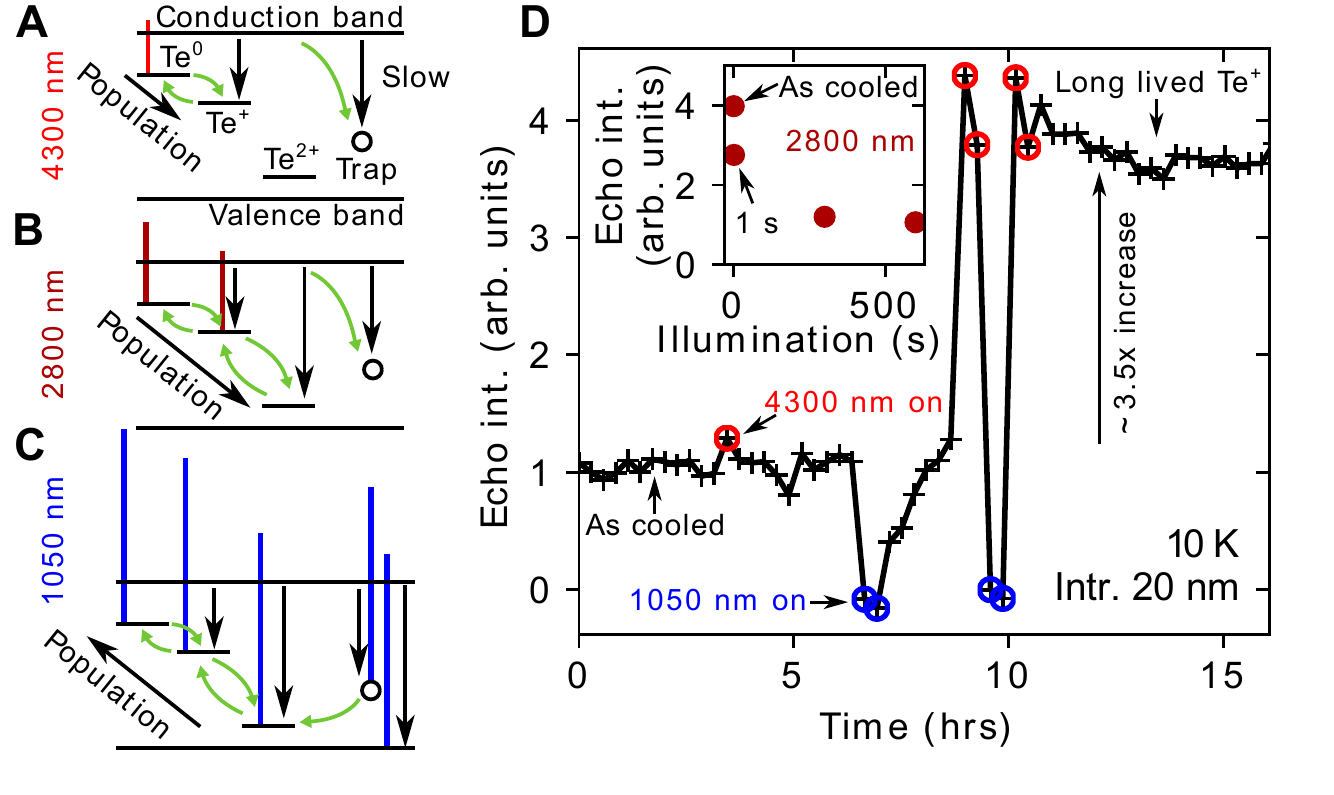}
    \caption{Energy levels and electron transitions of Te:Si under illumination of (\textbf{A}) 4300~nm, (\textbf{B}) 2800~nm and (\textbf{C}) 1050~nm light. The labels of energy levels are marked in (\textbf{A}). Coloured vertical lines show the energy of incident photons supplied by LEDs at cryogenic temperatures and indicate which levels are excited by these photons. Vertical arrows show electron capture. Green arrows indicate population swapping between levels under illumination with arrow direction indicating the shift of population of Te donors under illumination. (\textbf{D}) Echo intensity from the shallow-implanted intrinsic sample as a function of time as LEDs are shone at the sample during the echo acquisition (red and blue circles) at 10~K. 
    %With appropriate LED preparation, the echo intensity can be enhanced by a factor of $\sim 3.5\times$. 
    The inset shows the echo integral as a function of cumulative 2800~nm illumination showing that the echo is suppressed.}
    \label{fig:Optical}
\end{figure}

A comparison with the P:Si reference sample gives an estimated Te activation of $\sim 7\%$ for the shallow-implanted sample (Table~S1), which arises from a combination of imperfect incorporation of Te into the lattice and less than 100\% ionisation into Te$^+$.
%The charge configuration of donors can be modified by optical initialisation~\cite{nardo2015spin}.
Shining light of different wavelengths at this sample allows us to alter the charge configuration and improve the fraction of Te$^+$.
We explore illumination at three different wavelengths: 1050~nm, sufficient to excite carriers across the Si bandgap; 2800~nm, sufficient to promote electrons from Te$^+$ or Te$^0$ to the conduction band; and 4300~nm, which excites only the Te$^0$ state (see Fig.~\ref{fig:Optical}A-C).
Illumination at 2800~nm drives the formation of (spinless) Te$^{2+}$, as illustrated by the four-fold reduction of echo intensity as shown in the inset of Fig.~\ref{fig:Optical}D.

%We show experiments in Fig.~\ref{fig:Optical} where three infrared LEDs at 4300, 2800 and 1050~nm are shone at the sample at 10~K. We present a simple model for charge-state dynamics in Fig.~\ref{fig:Optical}(A-C), where light at 4300~nm (Fig.~\ref{fig:Optical}\textbf{A}) ionises Te$^0$ level forming Te$^+$ and a free electron. If that electron is recaptured by the Te$^+$, this leaves the Te charge configuration unchanged, but if the electron is captured by Te$^{2+}$ or some other trap, then this increases the Te$^+$ fraction. 
%Light at 2800~nm (Fig.~\ref{fig:Optical}B) causes the same Te$^0$ and Te$^+$ dynamics and can also ionise Te$^+$ forming Te$^{2+}$ which can in turn recapture the electron. If free electrons are trapped by a different centre in the silicon, then this illumination causes an accumulation of spinless Te$^{2+}$, reducing echo intensity, which is shown to be the case when plotting echo intensity vs. the integrated duration of 2800~nm illumination in the inset to Fig.~\ref{fig:Optical}D. 
%Light at 1050~nm is above the silicon band gap and ionises all states forming many free carriers and holes. The state after such illumination depends on the retrapping dynamics in the silicon. If Te$^+$ captures electrons faster than non-Te trap states in the silicon, then illumination will form Te$^0$ leaving empty traps. If the traps are faster than Te$^{2+}$, then the Te will remain doubly ionised after the traps rapidly capture electrons. 

In Fig.~\ref{fig:Optical}D we explore the effects of 4300~nm and 1050~nm illumination, with respect to the initial echo intensity from the sample as cooled from room-temperature to 10~K. Illumination first with 4300~nm results in a small ($\sim20\%$) increase in echo intensity, indicating there is only a small concentration of Te$^0$ in the as-cooled state, consistent with our simulations above.
Subsequent illumination at 1050~nm results in a complete suppression of the electron spin echo, consistent with driving population from Te$^+$ (and Te$^{2+}$) into the neutral Te$^0$ state. The echo recovered on the timescale of an hour, due to some redistribution of population from Te$^0$ to Te$^+$, similar to Se$^+$ recovery after illumination by 1047~nm~\cite{nardo2015spin}. However, as is evident from the large ($3.5\times$) increase in echo intensity following subsequent illumination at  4300~nm, there remained a substantial fraction of Te$^0$, much greater than that present upon cooling the sample. 
%
%Illuminating at 4300~nm LED after the dose of 1050~nm light gave a significant (. According to our model, this means that there was significant Te$^0$ present after the 1050~nm illumination and so Te$^+$ captures electrons much faster than non-Te traps. 
Further rounds of illumination at 1050 and 4300~nm demonstrate the ability to switch between Te$^0$ and Te$^+$ states. The non-equilibrium Te$^+$ population created persists for at least $\sim$16~hours (further data off panel) and represents a spin-activation fraction of $\sim26\%$. 
Similar illumination experiments applied to the deep-implanted p-type sample are described in the Supplementary Material, however, there no increase in echo intensity was observed relative to the as-cooled state.
The linewidths (half width at half maximum) of the \sx~and \sz~transitions of the deep-implanted p-type sample largely follow \dfdb, as shown in Fig.~S6A, consistent with inhomogeneous broadening from \sitwonine\ nuclear spins, as is commonly seen for donors in natural silicon. 
The linewidth reaches a minimum value of $\sim$0.6~MHz (Fig.~S6B), which is close to the pulse bandwidth limit ($\pi$-pulse duration 140~ns), but approximately equal to that measured for a clock transition in Bi:\natsi~doped at similar concentration~\cite{o2020spin}. The lineshape is well fit by a single Gaussian (see Fig.~S6B) with no evidence of additional splitting caused by isotope mass variation of the nearest-neighbour silicon atoms~\cite{sekiguchi2014host} --- the low ($I=1/2$) nuclear spin of $^{125}$Te means that the spin transitions are typically less sensitive to shifts in the hyperfine coupling than donors with high-spin nuclei like $^{209}$Bi. The increase in line broadening for \dfdb~$ \lesssim 0.03$ (corresponding to $B_0\lesssim 4$~mT) is due to SHF transitions splitting from the main transition as zero magnetic field is approached (see \SI\ for details).

Next we studied the spin relaxation time, $T_1$, and spin coherence time, \ttwo, in the temperature range 6.5--18~K for an \sx~transition at $9.65$~GHz and close to the \sz~clock transition (see Figs.~S10 and S11). In both cases, we observed the $T_1\propto T^{-9}$ temperature dependence indicating a (phonon-induced) Raman spin-relaxation process, as was also observed for $^{77}$Se$^+$ in \sitwoeight~\cite{nardo2015spin}. 
For temperatures above about 10~K, \ttwo~is limited by $T_1$, however, below 8~K, \ttwo~reaches a constant value. 

\begin{figure}
    \includegraphics[width=\linewidth]{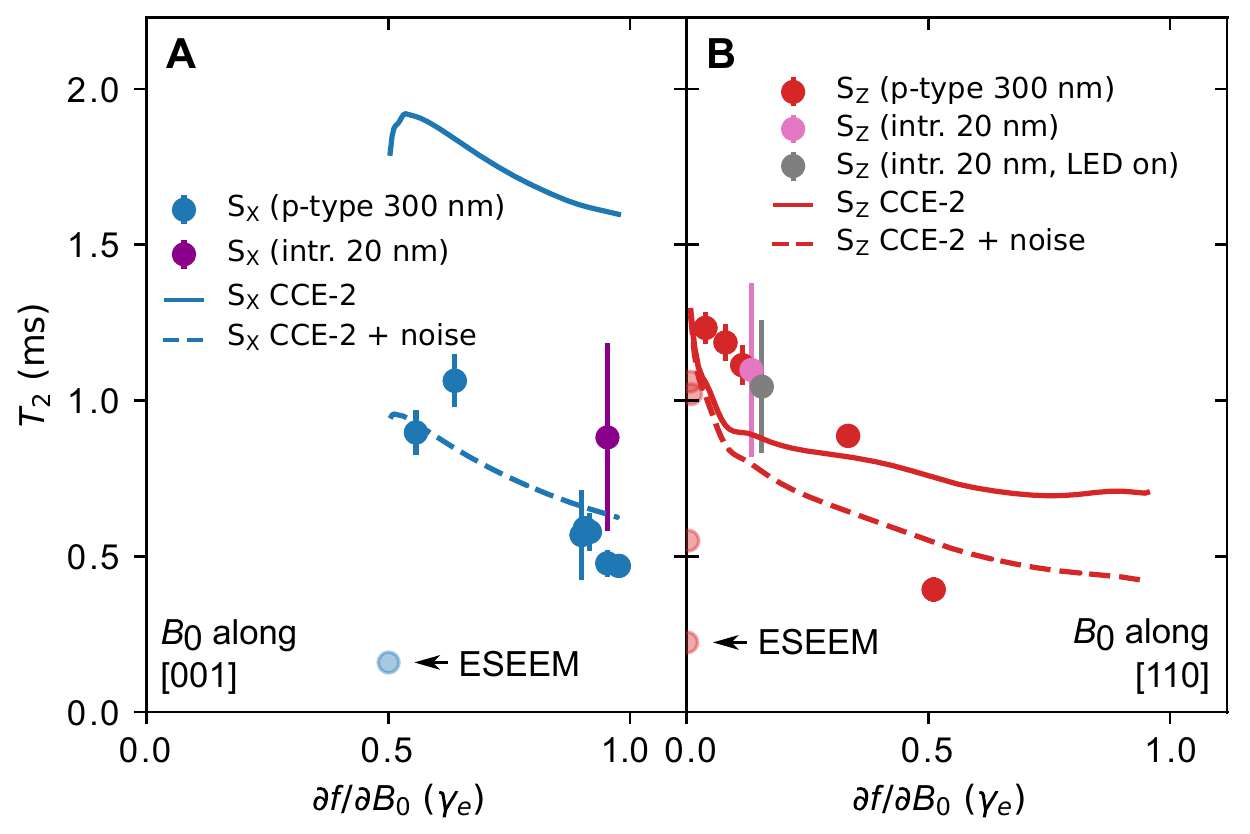}
    \caption{%(\textbf{A}) Temperature dependence of $T_1$ and \ttwo~times of $^{125}$Te$^+$ in deeply implanted p-type sample. Measurements performed at $9.646274$~GHz, $267.96$~mT (\sx) and $3.499200$~GHz, $4.90$~mT (\sz). $T_1$ data for \sz~transition are fit to a $T^{-9}$ dependence modelling a Raman relaxation process (green dashed line). The fit of the $T_1$ data of the \sx~transition is not shown for clarity. 
    \ttwo~as a function of \dfdb~for (\textbf{A}) \sx\ and (\textbf{B}) \sz\ transitions and different measurement conditions and samples. %The dashed curves in (\textbf{B}) represent a $B_0$-dependent noise model to the \ttwo~data. 
    Points where \ttwo~is suppressed by ESEEM are indicated with increased transparency. %The field regime, where ESEEM suppresses \ttwo, is shown in grey in (\textbf{C}) and the inset.}
    The solid curves mark the limit to \ttwo\ considering nuclear spin dynamics as a function of electron spin \dfdb. Adding in the effects of classical magnetic field noise gives good agreement to experiment (dashed curves).
    }
    \label{fig:Coherence}
\end{figure}

We investigate \ttwo~in this low-temperature limit for both \sx~and \sz~transitions in more detail, examining the effect of \dfdb. The results are summarised in Fig.~\ref{fig:Coherence}, including results from both the shallow- and deep-implanted samples.
We can expect that the naturally abundant \sitwonine\ nuclear spins in these samples pose a limit on the measured coherence times, which can be calculated using the cluster correlation expansion (CCE) method~\cite{yang2008quantum}, as used for (shallower) group-V donors~\cite{witzel2010electron, George2010}. We plot this calculated limit (CCE-2, based on two-body correlations of bath spins) along with the experimental results in Fig.~\ref{fig:Coherence}, with further details of the CCE simulations provided in the \SI. 

From the \ttwo\ measurements and simulations, we make four  observations. First is that the \ttwo\ measured for the shallow-implanted Te$^+$ in intrinsic substrate (ionised by surface band bending) reaches above 1~ms and is the same as for the deep-implanted Te$^+$ in p-type substrate (ionised by the boron acceptors). This suggests that neither the effect of the compensation, nor proximity to the surface limit the measured \ttwo\ values (see \SI\ for details).

Secondly, the \ttwo\ values of the \sx\ transitions are longer compared to the \sz\ transitions at the same \dfdb. This difference occurs due to different sample orientation with respect to $B_0$ direction when measuring the \sx\ ($B_0$ along [001]) and \sz\ (along [110]) transitions~\cite{abe2010electron} (see \SI\ for details). %For \sx\ experiments, $B_0$ is along the [001] axis, while \sz\ measurements are performed when $B_0$ is along the [110] axis. The effect of spectral diffusion from the \sitwonine\ bath spins in natural silicon strongly depends on this orientation due to the anisotropy of the nuclear dipolar interaction~\cite{abe2010electron}, as discussed in detail in \SI\ and illustrated in Fig.~S13 and S14.

The third observation is that the \ttwo\ limit from the \sitwonine\ nuclear spin bath predicted by the CCE-2 calculations is almost a factor of two longer than that seen for the shallow donors such as Bi and P~\cite{witzel2010electron, George2010}. However, as a result, the nuclear spin bath alone does not appear to account for the observed \ttwo\ values (see solid curves in Fig. \ref{fig:Coherence}A).
We consider additional sources of decoherence such as unhealed damage from ion implantation and spin concentration effects in the \SI\ and conclude that they are unlikely to play a significant role in these measurements.
It is known that experimental setups similar to that used here suffer from magnetic field noise that typically limits the measured \ttwo\ to $\sim 1$~ms. This effect can be circumvented by performing single-shot measurements of echo magnitude~\cite{ross2019electron}, but this requires a large signal-to-noise ratio not available when measuring implanted samples using 3D cavities.
Such an additional magnetic-field-type decoherence process (rate $\propto$ \dfdb) with a corresponding \ttwo~=~1~ms for \dfdb~$=\gamma_{\rm e}$ is able to rather well reproduce the observed measurements for both transitions (see dashed curves in Fig.~\ref{fig:Coherence}). We note that the simulated values of \ttwo\ for the \sz\ transition appear below the experimental data points, and we assign this discrepancy to a small misalignment of our sample with respect to the magnetic field.

%A misalignment from the principal axis of $\sim15^\circ$ for the \sz~measurements (nominally made for $B_0$ oriented along [001]) would be sufficient to bring the CCE-2 simulations in line with the measured \ttwo\ data. Additional measurements consistent with such a misalignment are presented in the Supplementary Materials, including a study of the \sx~transition \ttwo\ as a function of $B_0$ orientation.

Our fourth observation concerns the \ttwo\ behaviour as \dfdb\ approaches zero in the \sz~transitions (Fig. \ref{fig:Coherence}B). A key factor which distinguishes these measurements from those performed on a clock transition of Bi donors in natural silicon~\cite{wolfowicz2013atomic} is that here \dfdb$\rightarrow0$ coincides with the static magnetic field $B_0$ approaching zero (see Fig. \ref{fig:Transitions}C). Therefore, rather than seeing the \ttwo\ towards $\sim100$~ms (as for Bi:$^{\rm nat}$Si), the CCE-2 calculations predict an (orientation-dependent) limit of 1--2.5~ms. This can be understood as a `melting' of the `frozen-core' of \sitwonine\ spins around the donor spin, leading to enhanced noise from the spin bath. Furthermore, our measurements reveal an additional decrease in the fitted \ttwo\ at the very lowest values of \dfdb\, corresponding to $B_0\lesssim 4$~mT, which we ascribe to periodic electron spin echo envelope modulation (ESEEM) from \sitwonine\ nuclei in the crystal matrix. Simulations of the ESEEM performed using EasySpin~\cite{Stoll2006} accounting for this are presented in the \SI, and the same effect has been observed recently in Bi:Si~\cite{Probst2020}. 

The measured coherence times of the near-surface $^{125}$Te$^{+}$ spins are about an order of magnitude longer compared to other spin centres at similar depths \cite{Wang2016,Fukuda2018,Bluvstein2019} with clear prospects of further improvement, as each of the limiting decoherence processes described above can be mitigated by moving to isotopically purified silicon \cite{nardo2015spin,ranjan2020multimode,ranjan2021spatially,Morse2017}. %It is likely that this approach will allow to reach \ttwo\ of order of 100~ms as observed for similar donor systems~\cite{nardo2015spin,ranjan2021spatially,Morse2017}. %The micro-resonators will enable enhanced the signal to noise ratio (and thus single-shot measurements) while also providing greater control over the sample alignment with respect to the microwave field. Isotopically enriched \sitwoeight\ material will mitigate the effects of the \sitwonine\ bath responsible for the decoherenced captured in the CCE-2 simulations, as well as the ESEEM seen at very low magnetic fields. %Indeed, measurements of another deep donor (Se$^+$) in \sitwoeight\ showed \ttwo$~\sim$80~ms,  far from any clock transition, indicating the strong potential to increase coherence times in this platform~\cite{nardo2015spin}. 
Use of \sitwoeight\ should also substantially reduce the linewidths by removing the broadening from unresolved SHF levels. A better sample alignment and single-shot measurements can be achieved using superconducting resonators patterned onto implanted silicon \cite{ranjan2020multimode,o2020spin}.

In addition, the large (3.5~GHz) zero-field splitting of $^{125}$Te$^{+}$ makes it suitable for use at low magnetic fields and thus compatible with field-intolerant systems such as superconducting qubits. The nuclear spin-half of $^{125}$Te gives this donor an attractive level structure which can be used, for example, in so-called `flip-flop' qubits~\cite{tosi2017silicon}, and which permits near-complete polarisation at dilution fridge temperatures, even at zero field. %This is extremely useful, for example in spin ensemble quantum memories, where coupling strength is determined by the ensemble polarisation.  
The significant second ionisation energy permits the placement of spin-active $^{125}$Te$^{+}$ very close to surfaces and interfaces, which is beneficial for achieving large inductive coupling to microwave circuits~\cite{ranjan2020electron}. All of these features, combined with the optical transitions of such donors~\cite{Morse2017}, open a host of potential applications in quantum frequency converters, quantum sensors~\cite{maze2008nanoscale}, and quantum memories.

In conclusion, our results show that $^{125}$Te$^+$ in silicon is a promising donor for use in quantum technology applications. We have demonstrated spin coherence times in excess of 1~ms, for donors at depths of only 20~nm from the surface, and in an isotopically purified \sitwoeight\ substrate these may be expected to become even longer. We have also shown a novel approach to ionise shallowly implanted chalcogens in nanoelectronic devices using surface band bending, which, when combined with infrared illumination, gives a single ionisation fraction substantially greater than that achieved by co-doping with acceptors, and no visible reduction in coherence time in natural silicon.

This work was supported by the UK EPSRC Skills Hub in Quantum Systems Engineering: Innovation in Quantum Business, Applications, Technology and Engineering (InQuBATE), Grant No. EP/P510270/1; 
The European Research Council (ERC) via the LOQOMOTIONS grant (H2020- EU.1.1., Grant No. 771493).
R.B.L. was supported by Hong Kong Research Grants Council General Research Fund Project 14302121, and S.L. was supported by The Chinese University of Hong Kong Impact Postdoctoral Fellowship.
The authors acknowledge the UK National Ion Beam Centre (UKNIBC), where the silicon samples were ion implanted, and Nianhua Peng who performed the ion implantation.

%\section*{References}
\bibliography{bibliography}

%\section*{Author contributions}
%Conceptualization: M\v{S}, JO'S, OWK, RL, JJLM.
%Data curation: M\v{S}, JO'S, OWK, SL.
%Formal analysis: M\v{S}, JO'S, OWK, SL.
%Funding acquisition: M\v{S}, OWK, RL, JJLM.
%Investigation: M\v{S}, JO'S, OWK, SF, CWZ, SL.
%Methodology: M\v{S}, JO'S, OWK, RL, JJLM.
%Project administration: JJLM.
%Resources: M\v{S}, JO'S, GD, TS, PS, SF, RS, JJLM.
%Software: M\v{S}, JO'S, OWK, SL.
%Supervision: RL, JJLM.
%Validation: SF, RL, JJLM.
%Visualization: M\v{S}, JO'S, OWK.
%Writing – original draft: M\v{S}, JO'S, OWK, JJLM.
%Writing – review \& editing: all authors.

%\section*{Competing interests}
%The authors declare no competing interests.

\end{document}

\beginsupplement

%\begin{abstract}
%Superconducting qubits Quantum memories are a critical part of many 
%\end{abstract}

\title
{Supplementary Materials\\Near-surface $^{125}$Te$^+$ spins with millisecond coherence lifetime}

\author{Mantas~\v{S}im\.{e}nas}
\altaffiliation{These authors have contributed equally to this work} 
\affiliation{London Centre for Nanotechnology, UCL, 17-19 Gordon Street, London, WC1H 0AH, UK}

\author{James~O'Sullivan}
\altaffiliation{These authors have contributed equally to this work} 
\affiliation{London Centre for Nanotechnology, UCL, 17-19 Gordon Street, London, WC1H 0AH, UK}

\author{Oscar~W.~Kennedy}
\altaffiliation{These authors have contributed equally to this work} 
\affiliation{London Centre for Nanotechnology, UCL, 17-19 Gordon Street, London, WC1H 0AH, UK}

\author{Sen Lin}
\affiliation{Department of Physics, Centre for Quantum coherence and The Hong Kong Institute of Quantum Information Science and Technology, The Chinese University of Hong Kong, Hong Kong, China}

\author{Sarah Fearn}
\affiliation{Department of Materials, Imperial College London, London SW7 2BX, UK}

\author{Gavin Dold}
\affiliation{London Centre for Nanotechnology, UCL, 17-19 Gordon Street, London, WC1H 0AH, UK}

\author{Christoph~W.~Zollitsch}
\affiliation{London Centre for Nanotechnology, UCL, 17-19 Gordon Street, London, WC1H 0AH, UK}

\author{Tobias Schmitt}
\affiliation{Institute for Semiconductor Nanoelectronics, Peter Grünberg Institute 9, Forschungszentrum Jülich and RWTH Aachen University, Germany}

\author{Peter Sch\"uffelgen}
\affiliation{Institute for Semiconductor Nanoelectronics, Peter Grünberg Institute 9, Forschungszentrum Jülich and RWTH Aachen University, Germany}

\author{Ren-Bao Liu}
\affiliation{Department of Physics, Centre for Quantum coherence and The Hong Kong Institute of Quantum Information Science and Technology, The Chinese University of Hong Kong, Hong Kong, China}

\author{John~J.~L.~Morton}
\altaffiliation{jjl.morton@ucl.ac.uk} 
\affiliation{London Centre for Nanotechnology, UCL, 17-19 Gordon Street, London, WC1H 0AH, UK}
\affiliation{Department of Electrical and Electronic Engineering, UCL, Malet Place, London, WC1E 7JE, UK}

\maketitle

%\section*{Supplementary information}
\subsection{Samples}
We implanted mass-selected $^{125}$Te ions into three samples of 275~$\upmu$m thick float-zone silicon: two samples were p-type doped with [B]~$\sim 2.4(2)\times 10^{16}$~cm$^{-3}$, and one was intrinsic. The intrinsic sample and one p-type sample were ion implanted with $^{125}$Te at 20~keV with a fluence of 1.4$\times 10^{11}$~cm$^{-2}$ forming a shallow layer of dopants centred at a depth of $\sim$20~nm. The second p-type sample was implanted using 800~keV with a fluence 2.2$\times 10^{12}$~cm$^{-2}$ forming a $\sim200$~nm thick layer of Te centred at a depth of $\sim$300~nm. Monte Carlo simulations of ion implantation were performed using SRIM~\cite{ziegler2010srim} and are shown below alongside SIMS depth profiles before and after annealing. To activate the dopants, samples were annealed at 1000$^\circ$C for 5 minutes, unless otherwise stated. %For the remainder of this article, unless stated otherwise, we use the deeply implanted sample annealed at 1000$^\circ$C.

\subsection{SIMS experiments}
SIMS analysis of the samples was carried out using an IONTOF ToF-SIMS 5 instrument. The analytical ion beam used was a 25~keV Bi$_1^+$ LMIG. To obtain high resolution mass spectra and good detection limits, high current bunch mode was used with a beam current of about 1~pA. The analytical area was 100~$\upmu$m$^2$. Depth profiling was performed using a second sputtering Cs ion beam at 500~eV. The ion beam current was about 40~nA, and the sputter area was 300~$\upmu$m$^2$. Negative secondary ions were collected.

\subsection{Implanting and diffusion}
\begin{figure}[ht!]
    \centering
    \includegraphics[width=0.5\linewidth]{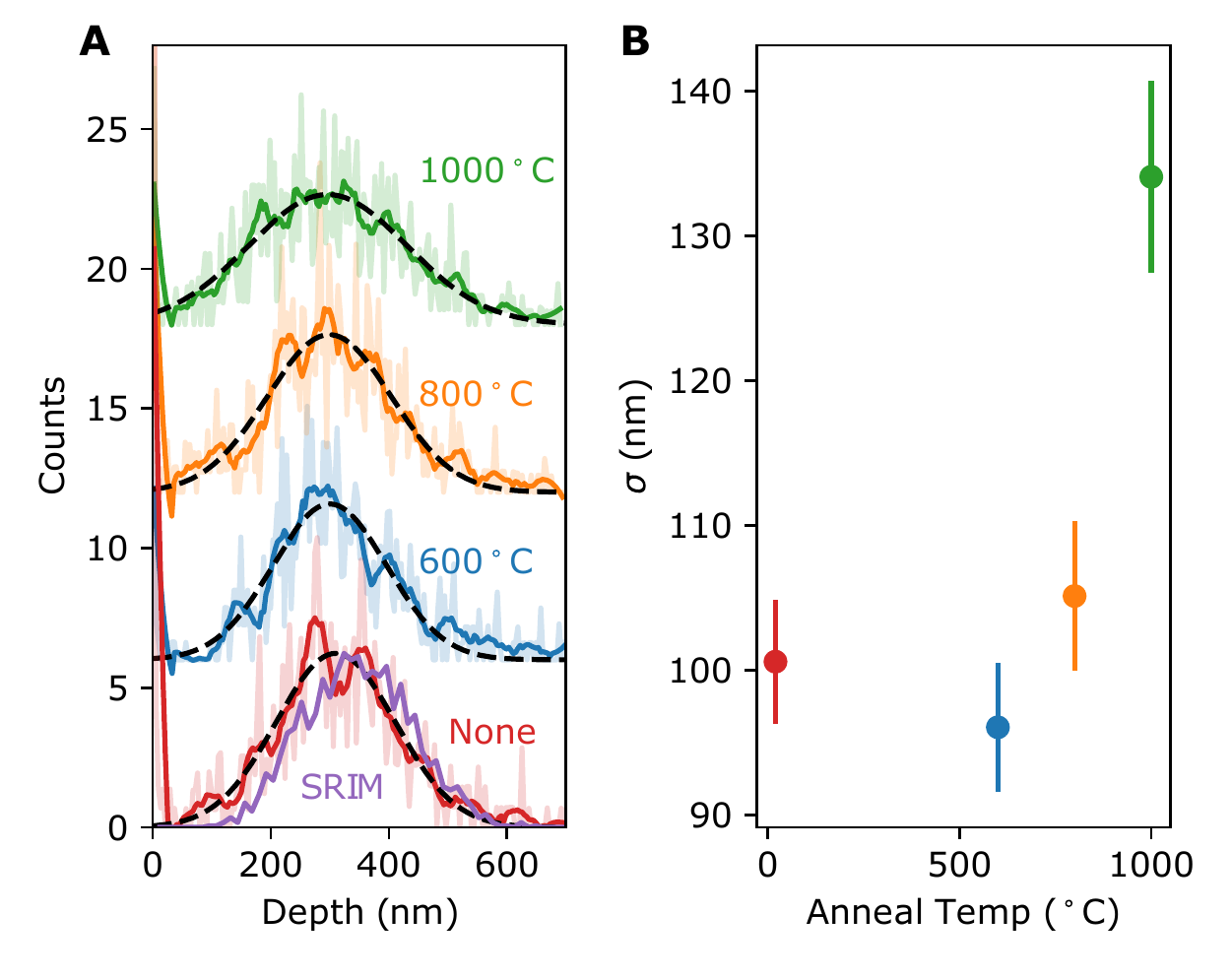}
    \caption{(\textbf{A}) $^{125}$Te counts in SIMS data of deep-implanted p-type sample (800~keV) normalised to Si$^{3-}$ counts to account for varying ion current. The faint curves are the measured data, while solid curves show the data after application of a Savitzky-Golay filter to reduce noise. Gaussians (black dashed lines) are fit to the raw data, and their standard deviation is shown in (\textbf{B}). The simulated Te distribution using SRIM is shown in (\textbf{A}). Slight differences in the centre of the peak may be caused by the depth calibration of SIMS data.}
    \label{fig:SIMS}
\end{figure}

\begin{figure}
    \centering
    \includegraphics[width=0.5\linewidth]{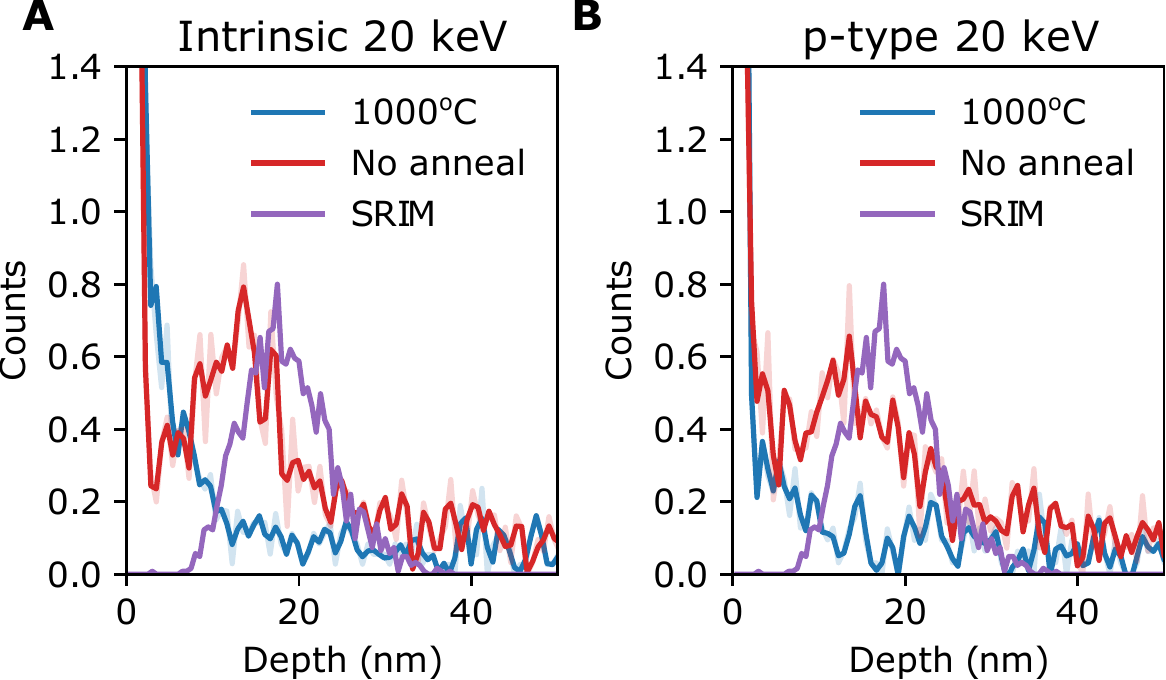}
    \caption{SIMS and SRIM for the (\textbf{A}) intrinsic and (\textbf{B}) p-type samples implanted at 20~keV before and after annealing at 1000$^\circ$C for 5 minutes.}
    \label{fig:SIMS20}
\end{figure}

We simulated the distribution of implanted ions using SRIM~\cite{ziegler2010srim}, a Monte-Carlo simulator, that simulates the random motion of energetic ions incident on samples. We show the SRIM profiles of Te for the 800~keV implantation in Fig.~\ref{fig:SIMS}A and the 20~keV implantation in Fig.~\ref{fig:SIMS20}. 

The SIMS profiles of samples before annealing match well to the SRIM profiles for both samples implanted at 20 and 800~keV (see Fig. \ref{fig:SIMS} and \ref{fig:SIMS20}). There is an offset in the depth from the SRIM and SIMS profiles with SRIM implying deeper implants than SIMS. The discrepancy is likely due to the non-equilibrium effects in the SIMS depth profiling when milling the surface. 

We measured the SIMS profile of Te after annealing the deep-implanted p-type sample at different temperatures (Fig.~\ref{fig:SIMS}A). The presented profiles are obtained by applying a Savitzky-Golay filter to smooth the noisy data. The Te distribution is well fit by a Gaussian profile for all annealing temperatures. We plot the standard deviation of the Gaussian profile as a function of annealing temperature in Fig.~\ref{fig:SIMS}B. There is little broadening of the implanted peak at temperatures $\lesssim 800^\circ$C, but the peak broadens by $\sim$35~nm at 1000$^\circ$C, showing that this annealing schedule allows reasonable diffusion in the sample. In the future, careful simultaneous optimization of donor diffusion, donor \ttwo~and activation yield will provide optimal anneal conditions.

We also measured the SIMS profiles for the samples implanted at 20~keV after annealing and show this data alongside SRIM simulations in Fig.~\ref{fig:SIMS20}. The peak in Te counts that was present before annealing is absent after annealing at 1000$^\circ$C. This is consistent with Te diffusion seen in the 800~keV sample. 

\subsection*{ESR experiments and analysis}
Pulsed ESR measurements at X-band ($\sim9.6$~GHz) were performed using a Bruker ELEXSYS E580 spectrometer equipped with a 1~kW travelling wave tube power amplifier and an overcoupled sapphire ring resonator (measured in reflection), resulting in a typical $\pi$-pulse duration of $\sim50$~ns. For measurements at S-band ($\sim3.5$~GHz), we used a home-built spectrometer (similar in design to that described in Ref.~[\citenum{o2020spin}]) and 3.5--4.1~GHz loop-gap resonators, wire abraded using oxygen-free copper, which were measured in transmission (see below for details). %The loop-gap resonators had as-fabricated quality factors of $\sim200$, reducing to $\sim50$, when the frequency was tuned by inserting a small (approximately 2x4~mm$^2$) piece of intrinsic 275~$\upmu$m float zone Si wafer into the capacitive gap, giving a window of about 100~MHz for experiments at different microwave frequencies. A typical $\pi$-pulse duration was $\sim700$~ns, when using a 30~W solid-state power amplifier. 
To amplify the weak spin echo signals from ion implanted samples, we used probeheads with cryogenic low-noise preamplifiers as described in Ref.~[\citenum{vsimenas2020sensitivity}]. Static magnetic fields were applied along [001] and [110] directions of the Si wafer for \sx\ and \sz\ geometries, respectively.

The EDFS spectra were recorded using Hahn echo pulse sequence. For fields below 4~mT, we fit the \sz\ EDFS spectra using the spin Hamiltonian to determine the centres of Gaussians with contributions from all $^{28}$Si nearest neighbours and also use the (111)$_2$ SHF transition leaving the amplitude and width of peaks as free parameters. The relaxation time \tone\ was obtained by fitting a monoexponential function to the data obtained by the inversion recovery sequence. The electron spin coherence time \ttwo\ was measured using the Hahn echo sequence. The resulting traces were fit using stretched-exponential decays to extract \ttwo\ times.

\subsection{S-band setup}
\begin{figure}
    \centering
    \includegraphics[width=0.5\linewidth]{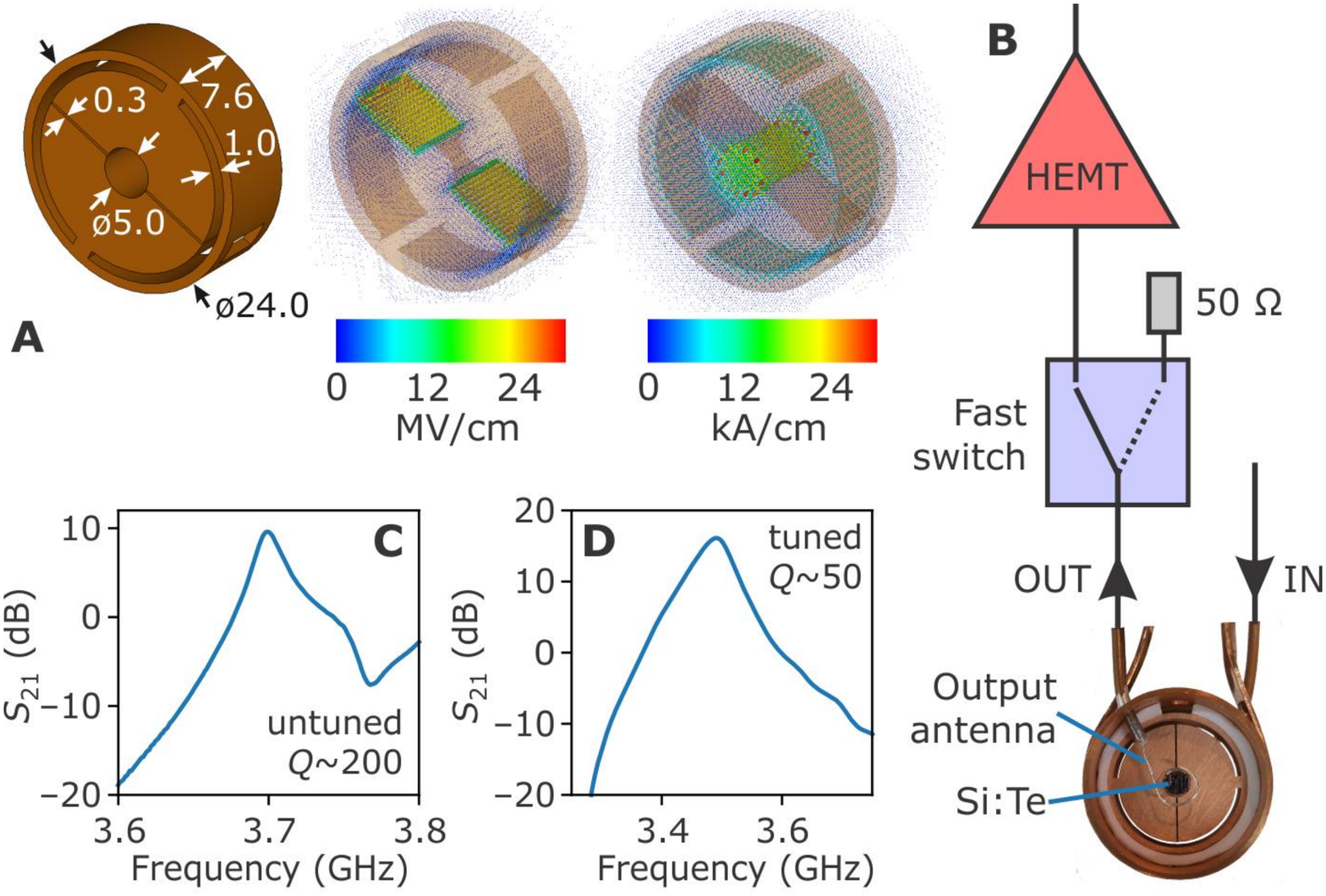}
    \caption{(\textbf{A}) Schematics and typical dimensions (in mm) of a loop-gap resonator, and simulation of the electric and magnetic microwave field components of a mode used to drive ESR transitions at S-band. Simulations are performed at arbitrary excitation power using CST Studio software. (\textbf{B}) Schematic of the microwave circuit used to amplify the spin signals. A photograph of a loop-gap resonator with inserted Te:Si sample is shown at the bottom in (\textbf{B}). The fast microwave switch is used to protect the low-noise cryogenic HEMT preamplifier from high power pulses. (\textbf{C},\textbf{D}) $S_{21}$ traces of untuned and tuned resonator.}
    \label{fig:setup}
\end{figure}

ESR experiments at S-band were performed using a home-built spectrometer \cite{o2020spin} with a 30~W solid-state power amplifier. The spectrometer was connected to a home-built transmission probehead equipped with a loop-gap resonator, wire abraded using oxygen-free copper. The resonator design is presented in Fig.~\ref{fig:setup}A together with the microwave field simulations of the mode used to drive the ESR transitions. The probehead contained a low-noise cryogenic high-electron-mobility transistor (HEMT) preamplifier (Low Noise Factory LNF-LNC6\_20C), which is protected from high power microwave pulses by a fast microwave switch (Analog Devices HMC547ALP3E) (see Fig. \ref{fig:setup}B). The resonator was coupled via a short input and a long output antenna and had as-fabricated quality factors of $\sim200$ (Fig. \ref{fig:setup}C), reducing to $\sim50$, when the frequency was tuned by inserting a small (approximately 2x4~mm$^2$) piece of intrinsic 275~$\upmu$m float zone Si wafer into the capacitive gap, giving a window of about 100~MHz for experiments at different microwave frequencies (Fig. \ref{fig:setup}D). A typical $\pi$-pulse duration was $\sim700$~ns, when using a 30~W solid-state power amplifier. Several resonators having different untuned frequencies and similar Q-factors were used in this study to cover $\sim$3.5--4.1 GHz frequency range.

\subsection{Spin Hamiltonian}
The singly ionised $^{125}$Te donor has a single bound electron ($S=1/2$) coupled to the $^{125}$Te nuclear spin ($I=1/2$) via an isotropic hyperfine coupling $A$ and is described by the following spin Hamiltonian in  frequency units:
\begin{equation}
\label{eq:ESR}
\mathbf{H}= A\mathbf{S\cdot I}+\gamma_{\rm e}\mathbf{B_0\cdot S}+\gamma_{\rm Te}\mathbf{B_0\cdot I}+\mathbf{H}_{\mathrm{Si}}. 
\end{equation}
Here, $\mathbf{B_0}$ is the applied magnetic field, while $\gamma_{\rm e}/2\pi = 28.025(1)$~GHz/T, $\gamma_{\rm Te}/2\pi = -13.5$~MHz/T  are, respectively, the gyromagnetic ratios of the bound electron and $^{125}$Te nucleus. The last term above describes the superhyperfine (SHF) interactions between the electron spin and neighbouring \sitwonine\ nuclei (natural abundance $\sim 4.7$\%):
\begin{equation}
\label{eq:shf}
\mathbf{H}_{\mathrm{Si}}= \sum_i{(\mathbf{S}\cdot \mathbf{A}_{\mathrm{ Si},i}+\gamma_{\rm Si}\mathbf{B_0}) \cdot \mathbf{I}_{\mathrm{Si},i}}.  
\end{equation}
Here, $\mathbf{A}_{\mathrm{ Si},i}$ is the SHF coupling to the $i$-th \sitwonine\ nuclear spin, $\mathbf{I}_{\rm Si}$, with gyromagnetic ratio $\gamma_{\rm Si}/2\pi = -8.5$~MHz/T.

\subsection{Te activation}
To determine the activated Te fraction, we measured the echo strength of the implanted samples relative to a reference sample. Both the reference sample and the implanted samples were together loaded into the cavity. Experiments were performed at X-band with pulse bandwidth larger than the spin linewidth and pulse parameters optimized for the cavity. The echoes were acquired in the centre of the respective spin transitions. The reference sample was a small piece of bulk doped P:Si containing a significant numbers of spins. The reference sample had resistivity $\rho = 2.2~\Omega\cdot$cm giving a phosphorous concentration of $2.1(2)\times 10^{15}$~cm$^{-3}$. Weighing the sample allowed us to determine the reference number of spins, $N_{\rm P}$.
To determine the number of implanted $^{125}$Te, $N_{\rm Te, total}$, we had to correct the implantation fluence to account for a non-perfect isotope selectivity of the implanter. The fractions of other implanted Te isotopes were obtained from the SIMS data. Then $N_{\rm Te, total}$ was determined by multiplying the surface area of the sample by the adjusted fluence. The surface area was determined from the mass of the used sample, the wafer thickness and silicon density.

We corrected the amplitudes of Te:Si and P:Si signals by the ratio of the matrix element squared for the two transitions, $M^2_{\rm rat} = M^2_{\rm P}/M^2_{\rm Te}$, and measured activation by comparing the echoes amplitudes from the two species denoted $A_{\rm Te}$ and $A_{\rm P}$, respectively. We computed the number of active Te as $N_{\rm Te, active} = N_{\rm P}\times M^2_{\rm rat}\times A_{\rm Te}/A_{\rm P}$ and found the activation by $N_{\rm Te, active}/N_{\rm Te, total}$.

The determined activation and number of spins $N_{\rm Te, active}$ of $^{125}$Te$^+$ in different samples are presented in Table~\ref{tab:activation}. Note that the number of spins in the shallow-implanted sample is about an order of magnitude lower compared to the deep-implanted p-type sample, resulting in much worse SNR, which significantly limits the choice of the ESR experiments. Further optimization of annealing schedules (e.g.\ longer anneals) would likely allow higher incorporation fraction, healing of implantation damage and control over diffusion. SIMS profiles also show some Te in the surface SiO$_2$, which is not spin active~\cite{Schenkel2009}, accounting for some reduction in the calculated activation.

\begin{table}[]
\begin{tabular}{|c|c|c|c|}
\hline
Sample                             & \ttwo~at X-band (ms) & Activation (\%) & $N_{\rm Te, active}/10^{11}$ \\ \hline
600$^\circ$C p-type 300~nm    & 0.24(2)                                  & 15(1)          & 2.02                         \\
800$^\circ$C p-type 300~nm    & 0.39(2)                                  & 18(2)          & 1.93                         \\
1000$^\circ$C p-type 300~nm   & 0.45(3)                                  & 22(2)           & 2.19                         \\
1000$^\circ$C intrinsic 20~nm & 0.44(9)                                  & 7.3(7)          & 0.193                        \\ \hline
\end{tabular}
\caption{\textbf{\ttwo~and activation of $^{125}$Te$^+$ in different samples.} The p-type and intrinsic samples were implanted at 800~keV and 20~keV, respectively. Activation and number of spins $N_{\rm Te, active}$ were determined by measuring the ratio of echo intensity between a reference P:Si sample and the Te-implanted samples.}
\label{tab:activation}
\end{table}

\subsection{LED illumination of the deep-implanted p-type sample}
\begin{figure}
    \centering
    \includegraphics[width=0.5\linewidth]{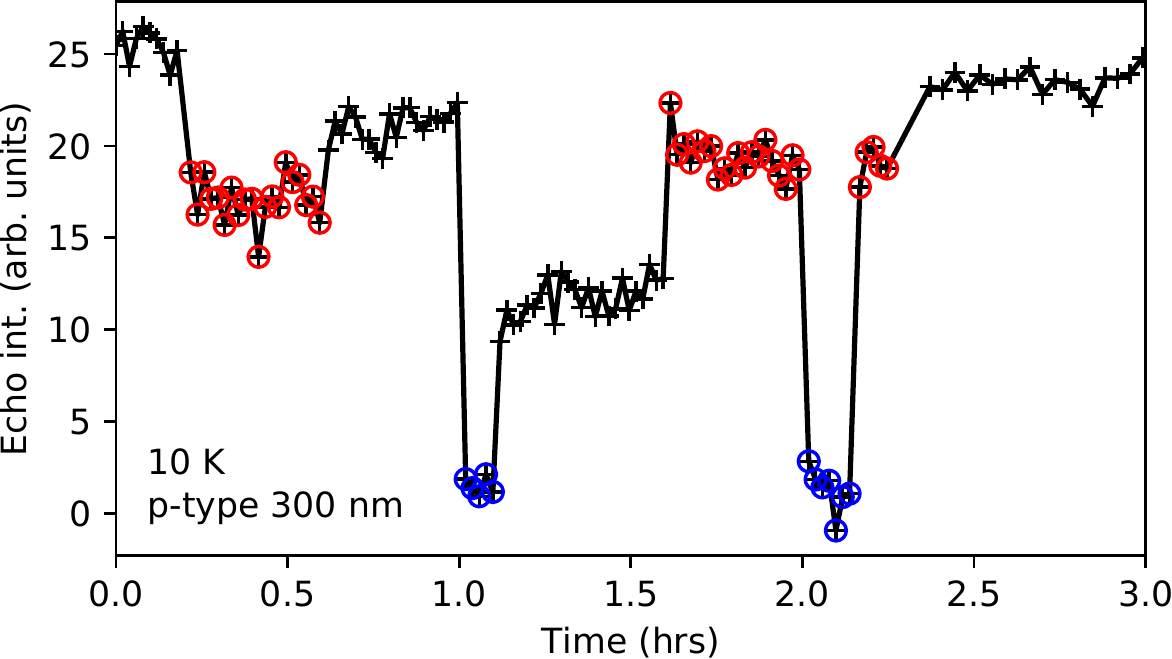}
    \caption{Echo intensity as a function of time as different wavelength light is shone at the deep-implanted p-type sample at 10~K. The sample starts as-cooled. Red and blue circles indicate 4300~nm and 1050~nm LED illumination, respectively, and black crosses are data without illumination.}
    \label{fig:p-typeLED}
\end{figure}

Fig. \ref{fig:p-typeLED} shows echo intensity as a function of time as different LEDs are shone at the deep-implanted p-type sample at 10 K. The 1050~nm illumination destroys the echo, while its intensity is recovered to the as-cooled value after the subsequent 4300~nm illumination.

% Boron: 22.4~$\Omega/\square$, 300~$\upmu$m thick -> 0.672~$\Omega$cm. Calculator gives 2.4$\times$10$^{16}$cm$^{-3}$. 

% We also investigate the coherence time for \sx~transitions at the X-band as a function of annealing temperature. We find much shorter coherence times for the sample annealed at 600$^\circ$C relative to those at 800$^\circ$C and 1000$^\circ$C as shown in Figure~\ref{fig:T2_anneal}. We attribute this change to improved healing of implantation damage by the higher temperature anneals. It may be that longer anneals at lower temperatures are also capable of healing implantation damage. 
% \begin{figure}
%     \centering
%     \includegraphics[width =\linewidth]{T2_anneal.pdf}
%     \caption{$T_2$ as a function of annealing temperature for the 800~keV p-type sample.}
%     \label{fig:T2_anneal}
% \end{figure}

\subsection{Fermi level pinning and band bending}
We used a 1D Poisson-Schr\"odinger solver which is well documented~\cite{tan1990self}. Our simulations defined the top surface as a Schottky barrier with 0~V applied across it, where the barrier height was varied, and the thick substrate boundary was an ohmic contact to ground. The top layer contained Te donors with the given concentrations.

We investigated FLP for different implantation profiles considering different bulk doping of the substrate. This allowed us to extract the band profile inside silicon, which determines the charge state of Te. We performed these simulations with a simulation temperature of 40~K, as the simulations do not converge at lower temperatures. Performing these simulations as a function of temperature shows that the band profiles depend very weakly on the simulation temperature.

\begin{figure}
    \centering
    \includegraphics[width=\textwidth]{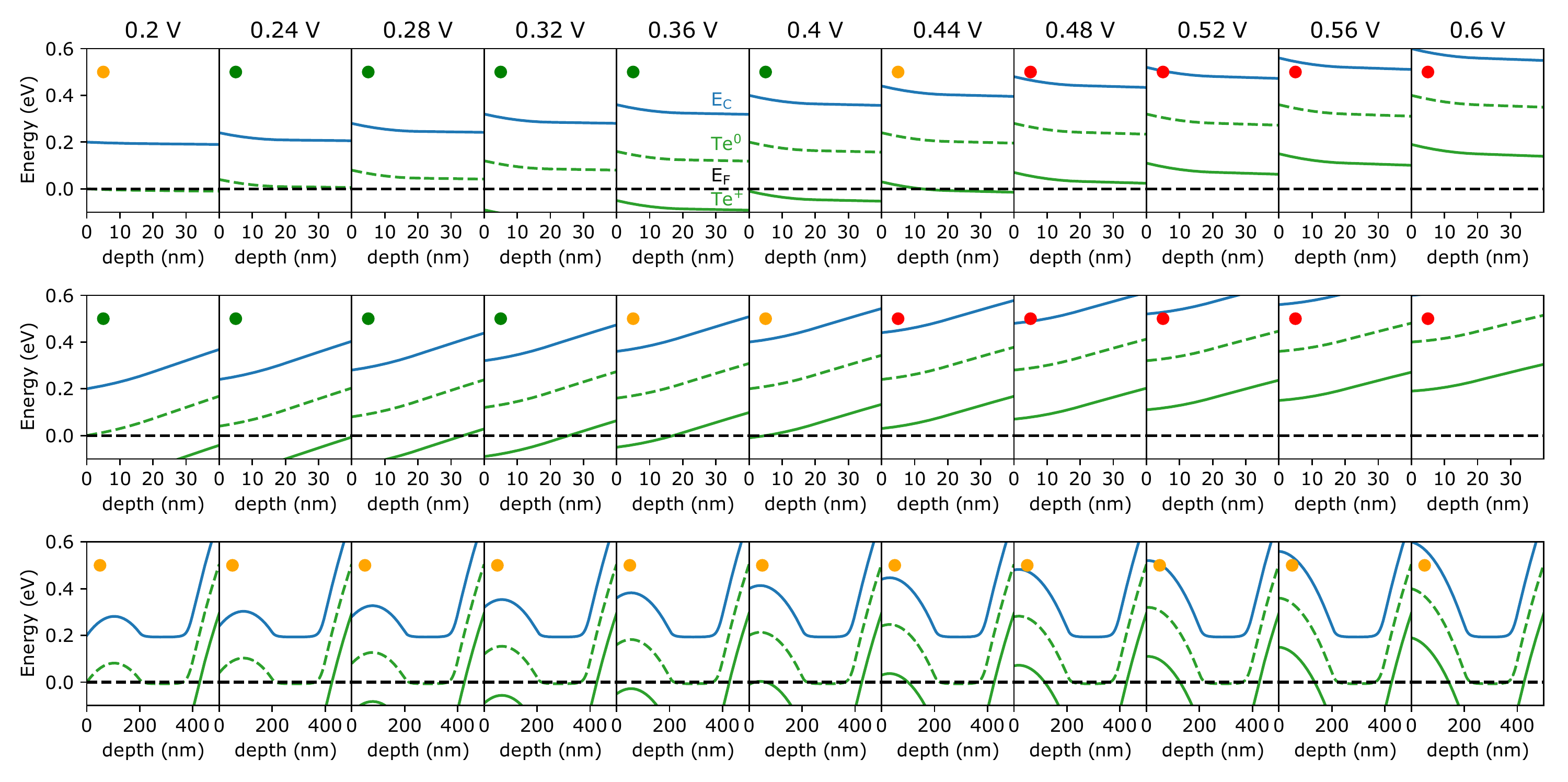}
    \caption{Effect of FLP voltage on the band structure at the surface in the Te doped samples. The top, middle and bottom rows correspond to 20~nm intrinsic, 20~nm p-type and 300~nm p-type samples, respectively. Each profile is colour coded as to whether it yields $\sim$100\% single ionisation (green), a fractional single ionisation (amber) or $\sim$0\% single ionisation (red).}
    \label{fig:FLP_Sims}
\end{figure}

We performed the simulations as a function of the FLP level, allowing us to investigate its effect on the ionisation fraction of Te. The results are shown in Fig.~\ref{fig:FLP_Sims}. We found regions of FLP, which result in high ionisation efficiency based on surface band bending for the shallow implants. In contrast, the deeper implant has approximately constant ionisation fraction independent of the surface and determined by the boron concentration.

\subsection{Linewidth and lineshape}
\begin{figure}
    \centering
    \includegraphics[width=0.65\textwidth]{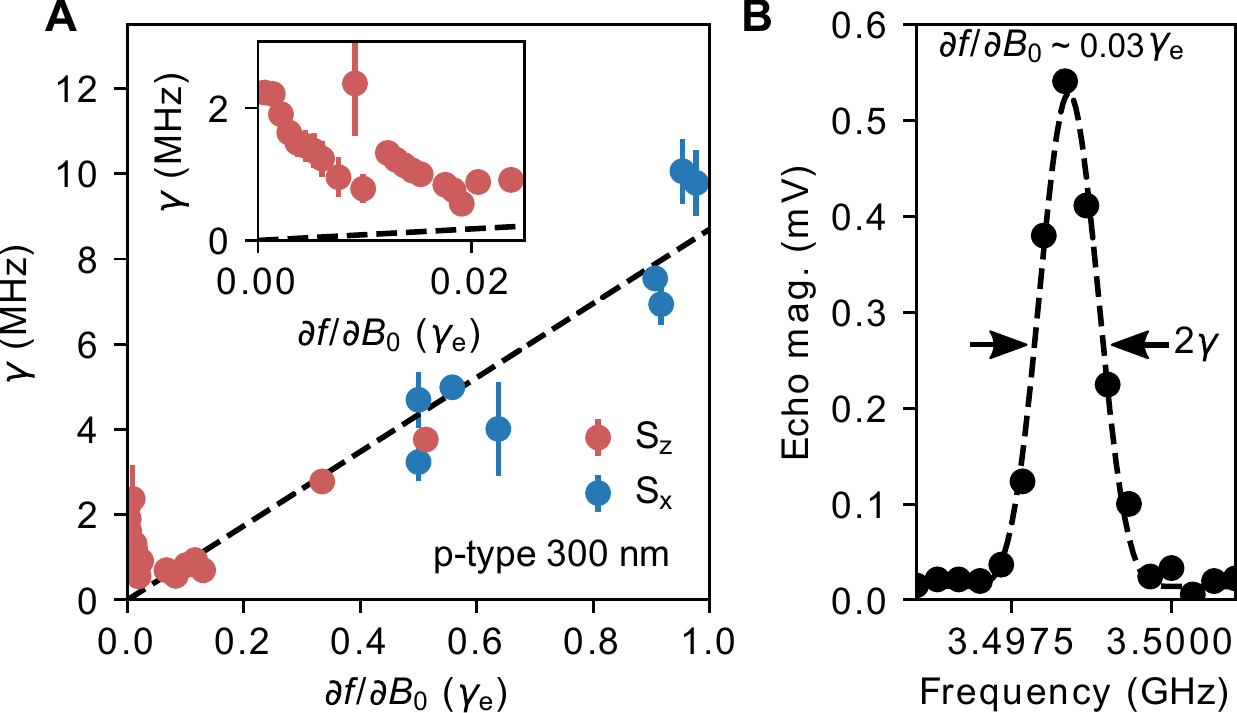}
    \caption{(\textbf{A}) Half width at half maximum linewidth of the \sx~(blue) and \sz~(red) transitions of deep-implanted p-type sample as a function of \dfdb~determined by fitting a Gaussian to the frequency-swept echo amplitude. Inset shows the increase in linewidth observed for \sz\ transition at low \dfdb\ arising from unresolved SHF transitions. 
    %when single Gaussians do not reproduce the lineshape, data have been omitted. 
    (\textbf{B}) The lineshape measured at the narrowest linewidth is fit well by a single Gaussian.}
    \label{fig:Linewidth}
\end{figure}

The ESR linewidths of the \sx~and \sz~transitions as a function of \dfdb~are shown in Fig.~\ref{fig:Linewidth}A. The lineshape measured at the narrowest linewidth ($\sim$0.6~MHz) is fit well by a Gaussian as shown in Fig.~\ref{fig:Linewidth}B.

\subsection{SHF levels}
\begin{figure}
    \centering
    \includegraphics[width=0.95\textwidth]{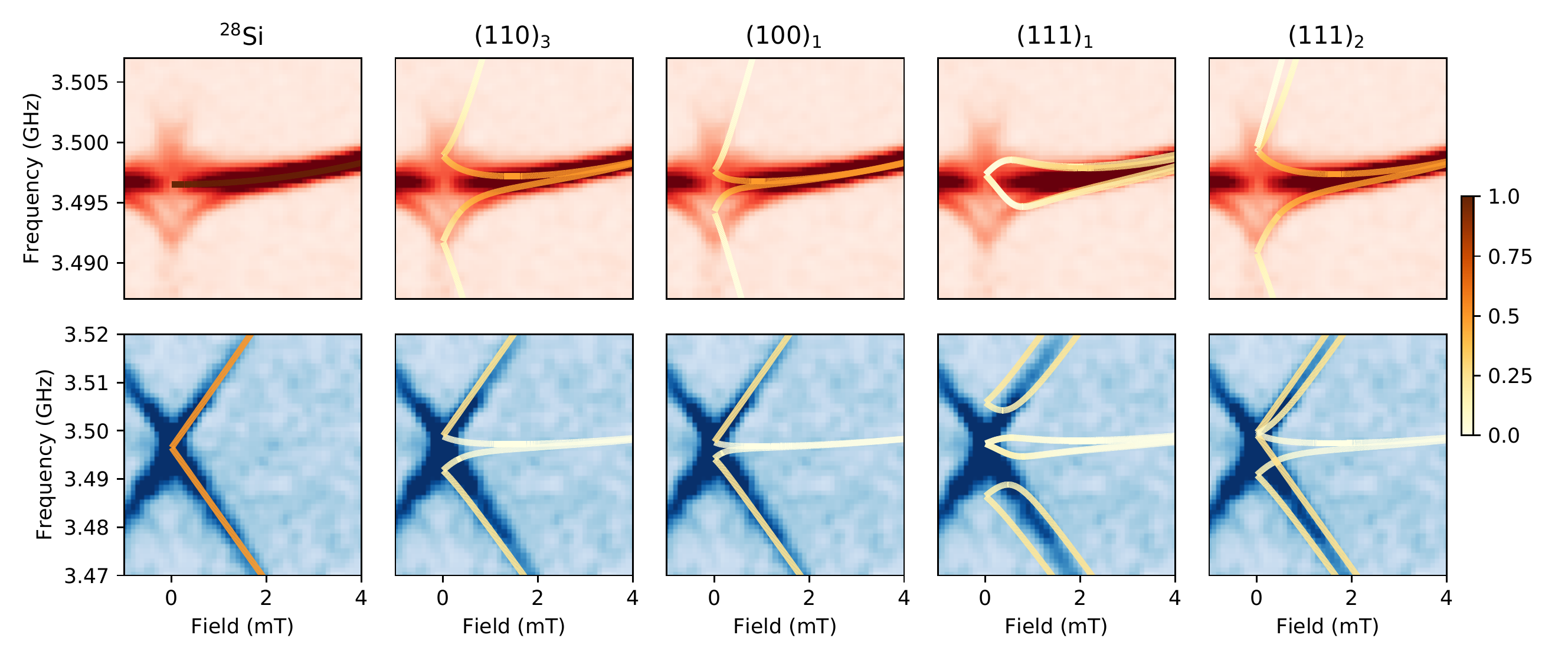}
    \caption{Different SHF transitions overlaid on the experimental 2D EDFS data of deep-implanted p-type sample (300~nm). The top (bottom) row shows \sz~(\sx)~transitions. The line colour of SHF levels is proportional to the matrix element squared. There are differences in the number of equivalent subsites for the different shells (see Ref.~\cite{niklas1983endor}) meaning that the intensity of the different SHF branches in EDFS will be modified.}
    \label{fig:SHF_levels}
\end{figure}

SHF effects shift the frequency of the $^{125}$Te$^+$:Si ESR transitions. The SHF coupling contains dipolar and contact terms and depends on the location of $^{29}$Si nuclear spins with respect to $^{125}$Te$^+$. ENDOR measurements of the coupling parameters for the closest $^{29}$Si have been reported in Ref.~\cite{niklas1983endor}. In Fig.~\ref{fig:SHF_levels}, we overlay experimental EDFS data with the simulated transitions corresponding to the strongest SHF couplings with lines colourised by the matrix element squared at the different transitions. Simulations were performed using EasySpin~\cite{Stoll2006}.

We can see that the \sz~transition of Te splits into well resolved levels, and that the different levels are well matched by the SHF levels from the $^{29}$Si considered here. The \sx~transition is sufficiently broad that the different SHF levels cannot be well resolved. However, for \sx~transitions there is asymmetry between the regimes higher and lower in frequency than the zero-field splitting, and the SHF levels are split further from the all-$^{28}$Si case at lower frequencies. This is reflected in the linewidth of the \sx~transition with $\gamma \sim 3$~MHz at lower frequencies and $\gamma \sim 4$~MHz at higher frequencies with both these points shown in Fig.~\ref{fig:Linewidth}.

\begin{figure}
    \centering
    \includegraphics[width=0.5\linewidth]{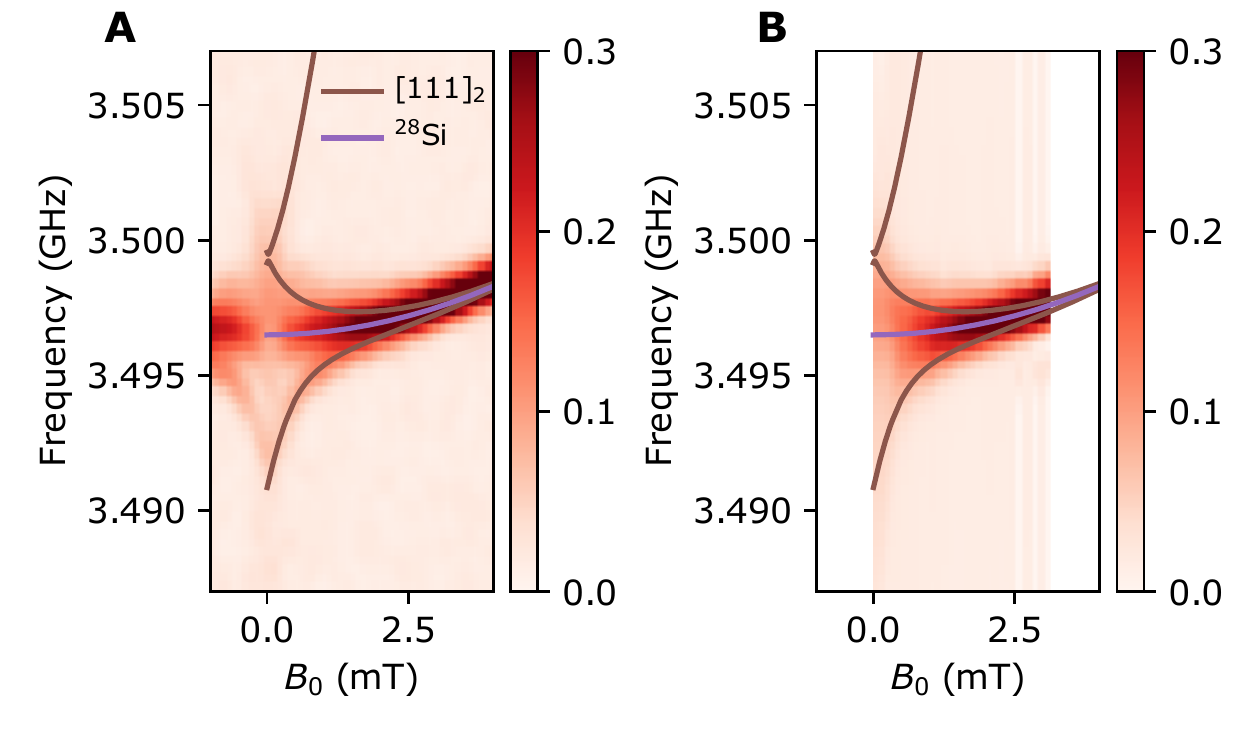}
    \caption{(\textbf{A}) Repeat of the EDFS from Fig. 1A (main text) showing the SHF levels used to fit the spectrum. (\textbf{B}) The result of the fit using four Gaussians centred at the lines shown in the figure. This fit is used to extract linewidths at low \dfdb~in Fig.~\ref{fig:Linewidth}.}
    \label{fig:Hamiltonian_EDFS}
\end{figure}

We fit the 2D-EDFS for \sz~transitions considering all-$^{28}$Si and the (111)$_2$ lines. The fit results are shown in Fig.~\ref{fig:Hamiltonian_EDFS}.

\subsection{ESEEM}
\begin{figure}
    \centering
    \includegraphics[width=0.5\linewidth]{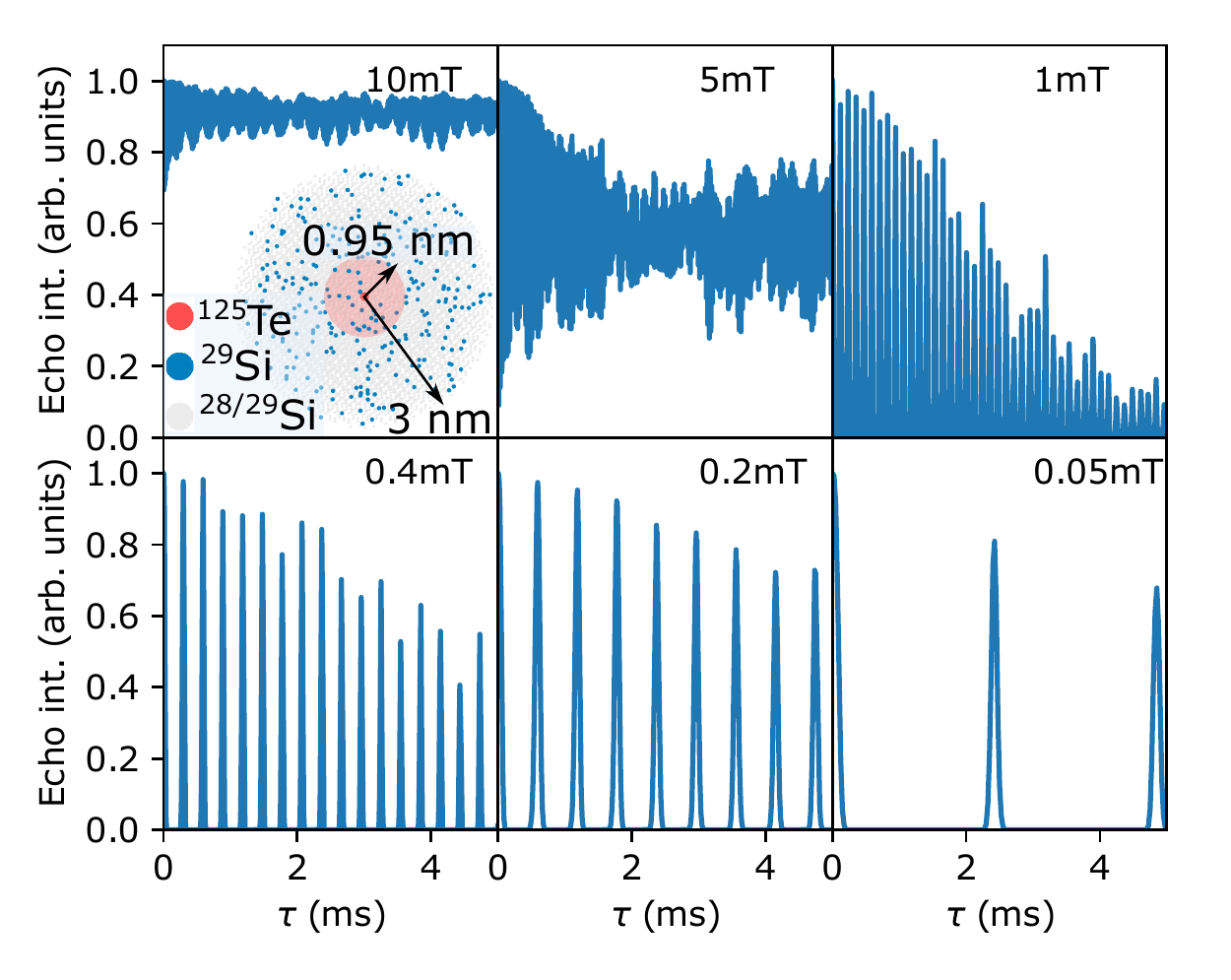}
    \caption{Two pulse ESEEM modulations at different field magnitudes. We model ESEEM modulation using the model shown in inset, where $^{29}$Si nuclei are randomly distributed in the matrix around the central $^{125}$Te atom. At higher fields, ESEEM modulation results in minimal change to echo amplitude, but at lower fields strong ESEEM results in a periodic and complete suppression of echo amplitude.}
    \label{fig:ESEEM_sim}
\end{figure}

We used EasySpin~\cite{Stoll2006} to simulate the echo suppression by ESEEM at low magnetic fields. In our simulation, $^{29}$Si was randomly distributed according to its natural abundance around $^{125}$Te$^+$ centre in Si lattice (see inset in Fig.~\ref{fig:ESEEM_sim}). For $^{125}$Te$^+$--$^{29}$Si distances of less than 9.5~\AA, we used the SHF couplings determined in the previous ENDOR study \cite{niklas1983endor}. For longer distances, we ignored the isotropic contact contribution and calculated the SHF couplings using the point-dipole approximation \cite{schweiger2001pulsedEPR}. The interaction distance was truncated at 30~\AA, which proved to be sufficient to achieve a good convergence of the simulation. We averaged 50 of such realizations to get the final ESEEM traces.

\begin{figure}
    \centering
    \includegraphics[width=0.5\linewidth]{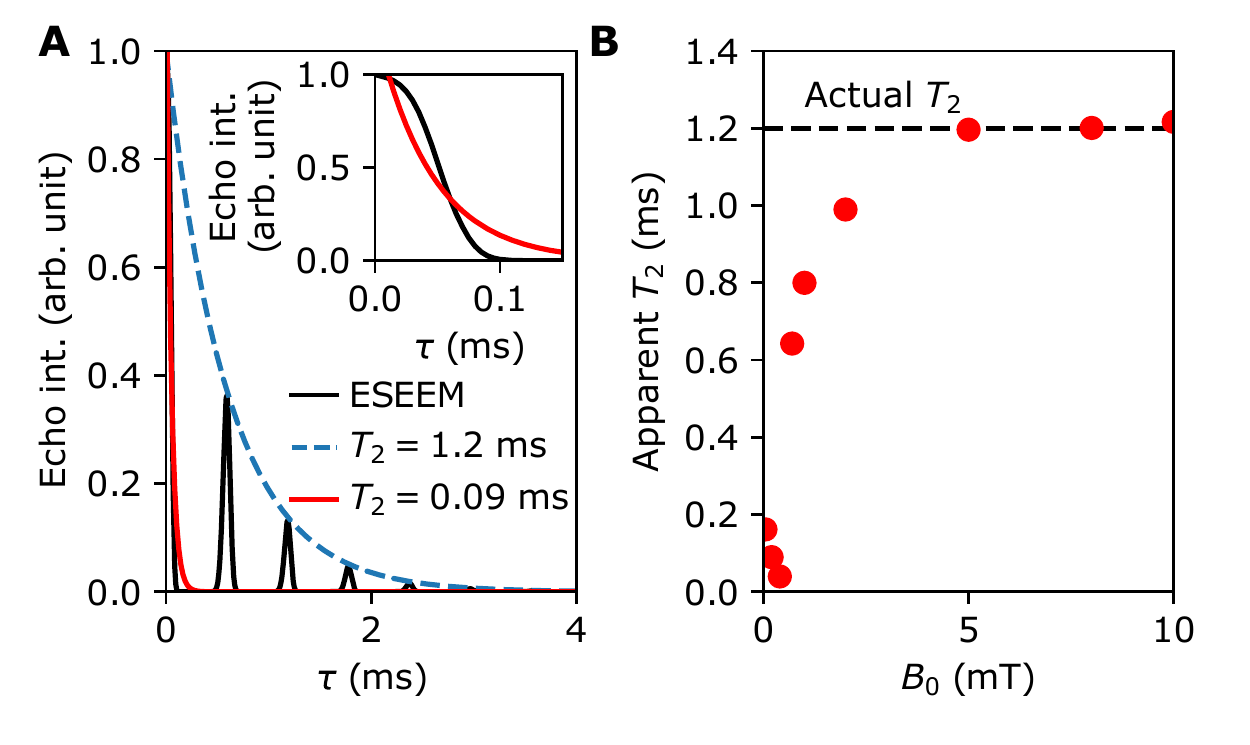}
    \caption{(\textbf{A}) Simulated coherence decay with ESEEM of $^{125}$Te$^+$:Si at 0.2~mT assuming $T_2=1.2$~ms (black). The exponential fit to the data (red) results in a much shorter apparent \ttwo~of 0.09~ms due to the severe ESEEM. Inset shows the initial part of the trace. (\textbf{B}) Magnetic field dependence of the apparent \ttwo~obtained by fitting the simulated coherence decays.}
    \label{fig:ESEEM_T2}
\end{figure}

The simulated time domain ESEEM data obtained at different magnetic fields are presented in Fig. \ref{fig:ESEEM_sim} revealing ESEEM with the modulation frequency close to the Larmor frequency of $^{29}$Si. This indicates that ESEEM is mostly dominated by very weakly coupled $^{29}$Si nuclei. At magnetic fields below $\sim 1$~mT, the time interval between the dips in the echo intensity becomes comparable with \ttwo~of $^{125}$Te$^+$ resulting in a much shorter value of the effective apparent \ttwo~below $\sim 5$~mT as demonstrated in Fig. \ref{fig:ESEEM_T2}. This result is in agreement with our experimental results that show a sudden decrease of the measured \ttwo~in the same field region.

\subsection{CCE simulations}
The spins have the interactions
\begin{equation}
H=H_{\mathrm{Te}}+H_{\mathrm{Si}}+H_{\mathrm{int}},  \label{H}
\end{equation}%
with
\begin{subequations}
\begin{eqnarray}
H_{\mathrm{Te}} &=&\omega _{\rm e}S_{z}+A_{0}\mathbf{S}\cdot \mathbf{I}%
_{0}+\omega _{\mathrm{n}}^{\mathrm{Te}}I_{0}^{z}, \\
H_{\mathrm{int}} &\approx &S_{z}\sum_{i}\hat{z}\cdot \mathbb{A}_{i}\cdot
\mathbf{I}_{i}\equiv S_{z}b_{z}, \\
H_{\mathrm{Si}} &=&\omega _{\mathrm{n}}^{\mathrm{Si}}\sum_{i}I_{i}^{z}+%
\sum_{i<j}\mathbf{I}_{i}\cdot \mathbb{D}_{ij}\cdot \mathbf{I}_{j},
\end{eqnarray}%
where $\mathbf{S}$ is the $^{125}$Te donor electron spin operator, $\mathbf{I}_{0}$ ($\mathbf{I}_{i}$) is the $^{125}$Te ($^{29}$Si) nuclear spin operator, $\omega _{\rm e}$, $\omega _{\mathrm{n}}^{\mathrm{Te}}$, and $\omega _{\mathrm{n}}^{\mathrm{Si}}$ are correspondingly the Larmor frequencies of the donor electron spin, donor nuclear spin and bath nuclear spin, which are related to their gyromagnetic ratios $\gamma $ by $\omega _{\alpha }=\gamma_{\alpha }B_{0}$, and therein $B_{0}$ is the external magnetic field applied along $z$ axis. $A_{0}$ ($\mathbb{A}_{i}$) denotes the hyperfine coupling strength (tensor) between the donor electron spin and the $^{125}$Te ($^{29}$Si) nuclear spins, and $\mathbb{D}_{ij}$ is the nuclear-nuclear dipole interaction tensor. Eq.~4b is valid in the limit $\omega_{\rm ee} \gg b$ (or equivalently the electron Zeeman energy is much larger than the hyperfine coupling of $^{125}$Te$^+$ to silicon). 

When there is no external field ($\omega _{\rm e}=\omega _{\mathrm{n}}^{\mathrm{%
Te}}=0$), the eigenstates of $H_{\mathrm{Te}}$ are the triplet states $%
\left\vert T_{+}\right\rangle =\left\vert \uparrow \uparrow \right\rangle
,\left\vert T_{0}\right\rangle =\left( \left\vert \uparrow \downarrow
\right\rangle +\left\vert \downarrow \uparrow \right\rangle \right) /\sqrt{2}%
,\left\vert T_{-}\right\rangle =\left\vert \downarrow \downarrow
\right\rangle $ with the eigenenergy $A_{0}/4$ and singlet states $%
\left\vert S_{0}\right\rangle =\left( \left\vert \uparrow \downarrow
\right\rangle -\left\vert \downarrow \uparrow \right\rangle \right) /\sqrt{2}
$ with $-3A_{0}/4$. The existence of magnetic field can mix the singlet
state $\left\vert S_{0}\right\rangle $ and the triplet state $\left\vert
T_{0}\right\rangle $ by the Zeeman energy that leads to two new eigenstates
\end{subequations}
\begin{eqnarray}
\left\vert u\right\rangle &=&\cos \frac{\theta }{2}\left\vert
T_{0}\right\rangle +\sin \frac{\theta }{2}\left\vert S_{0}\right\rangle \\
\left\vert d\right\rangle &=&-\sin \frac{\theta }{2}\left\vert
T_{0}\right\rangle +\cos \frac{\theta }{2}\left\vert S_{0}\right\rangle
\end{eqnarray}%
with corresponding eigenenergies%
\begin{equation}
E_{\mathrm{u}/\mathrm{d}}\approx -\frac{A_{0}}{4}\pm \frac{\Omega +b_{z}\sin
\theta }{2},
\end{equation}%
where $\theta =\arctan \left[ \omega _{\rm e}/A_{0}\right] $ and $\Omega =\sqrt{%
A_{0}^{2}+\omega _{\rm e}^{2}}$. Here, we have omitted $\omega _{\mathrm{n}}^{%
\mathrm{Te}}$ since it is much smaller than $\omega _{\rm e}$. The other two
eigenstates $\left\vert T_{\pm }\right\rangle $ remain unchanged, but their
eigenenergies become
\begin{equation}
E_{T_{\pm }}=\frac{A_{0}}{4}\pm \frac{\omega _{\rm e}+b_{z}}{2}.
\end{equation}

The Hamiltonian can be written as
\begin{equation*}
H\approx |1\rangle \langle 1|\otimes H^{(+)}+|0\rangle \langle 0|\otimes
H^{(-)},
\end{equation*}%
where the central-spin-conditional bath Hamiltonians for transition $%
\left\vert u\right\rangle \leftrightarrow \left\vert d\right\rangle $ ($S_{z}
$ transition) are given as
\begin{subequations}
\begin{eqnarray}
H^{\left( +\right) } &\approx &-\frac{A_{0}}{4}+\frac{\Omega }{2}+\frac{%
b_{z}\sin \theta }{2}+H_{\mathrm{Si}}, \\
H^{\left( -\right) } &\approx &-\frac{A_{0}}{4}-\frac{\Omega }{2}-\frac{%
b_{z}\sin \theta }{2}+H_{\mathrm{Si}},
\end{eqnarray}%
and for transition $\left\vert u\right\rangle \leftrightarrow \left\vert
T_{-}\right\rangle $ ($S_{x}$ transition)
\end{subequations}
\begin{subequations}
\begin{eqnarray}
H^{\left( +\right) } &\approx &-\frac{A_{0}}{4}+\frac{\Omega }{2}+\frac{%
b_{z}\sin \theta }{2}+H_{\mathrm{Si}}, \\
H^{\left( -\right) } &=&\frac{A_{0}}{4}-\frac{\omega _{\rm e}}{2}-\frac{b_{z}}{2}%
+H_{\mathrm{Si}}.
\end{eqnarray}

The Hahn echo signal is
\end{subequations}
\begin{equation}
A_{\rm e}\propto \mathcal{L}\left( 2\tau \right) =\mathrm{Tr}\left( \rho _{\mathrm{%
n}}e^{iH^{\left( -\right) }\tau }e^{iH^{\left( +\right) }\tau
}e^{-iH^{\left( -\right) }\tau }e^{-iH^{\left( +\right) }\tau }\right) ,
\notag
\end{equation}%
where $\rho _{\mathrm{n}}$ is the initial bath density matrix (which we choose as the infinite high-temperature thermalized state since the nuclear Zeeman energy is much less than 10~mK).

The central spin coherence was calculated using the CCE~\cite{yang2008quantum}, in which the decoherence caused by a cluster of $M$ bath spins $(1,2,3,M)$ is denoted as ${\mathcal{L}}_{1,2,\ldots ,M}$. The irreducible correlation of a cluster is defined recursively as $\tilde{%
\mathcal{L}}_{j}={\mathcal{L}}_{j}$, $\tilde{\mathcal{L}}_{i,j}\equiv {{%
\mathcal{L}}_{i,j}}{\tilde{\mathcal{L}}_{i}^{-1}\tilde{\mathcal{L}}_{j}^{-1}}
$, etc, that is, the decoherence function divided by all irreducible
correlations of all sub-clusters. For the $M$-order truncation (CCE-$M$),
the calculation takes into account the irreducible correlations up to the
clusters of $M$ spins, ${\mathcal{L}}\approx {\mathcal{L}}^{(M)}$, with
\begin{equation}
{\mathcal{L}}^{(M)}=\prod_{i_{1}}\tilde{\mathcal{L}}_{i_{1}}%
\prod_{j_{1}<j_{2}}\tilde{\mathcal{L}}_{j_{1},j_{2}}\cdots
\prod_{k_{1}<k_{2}\cdots <k_{M}}\tilde{\mathcal{L}}_{k_{1},k_{2},\ldots
,k_{M}}.  \notag
\end{equation}%
With the secular approximation, the CCE-1 contribution (decoherence due to single-spin dynamics, which also causes the ESEEM for relatively strongly coupled nuclear spins) vanishes.

In the simulation, we place the \sitwonine\ nuclear spins (with a natural abundance of 0.047) randomly on the Si lattice sites and the Te$^+$ ions randomly substituting one of Si. The bath includes all nuclear spins within a sphere of radius of 5~nm around the central spin. A larger bath size produces nearly the same result. The simulation is carried out by using an ensemble average over many (100) different spatial configurations. The convergence of CCE is confirmed by the fact that CCE-3 and CCE-2 produce nearly identical results.

\subsection{\tone~and \ttwo}
\begin{figure}[ht!]
    \centering
    \includegraphics[width=0.5\linewidth]{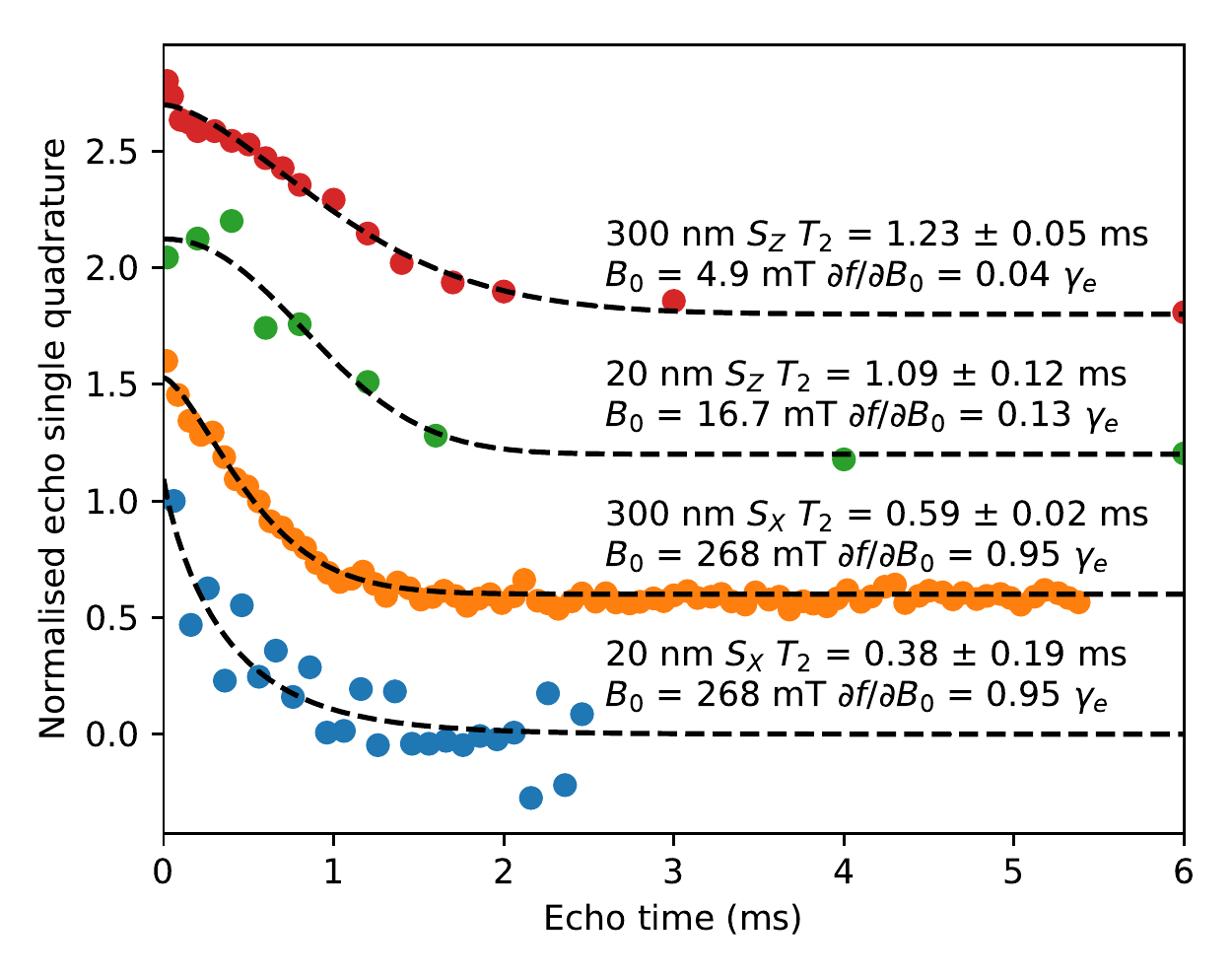}
    \caption{Selected \ttwo~decays of \sx~and\sz~transitions of deep- and shallow-implanted samples. The dashed curves correspond to the stretched exponential fits.}
    \label{fig:T2_egfits}
\end{figure}

The selected \ttwo~decays for the deep- and shallow-implanted samples for both \sx\ and \sz\ transitions are presented in Fig. \ref{fig:T2_egfits}.

\begin{figure}
    \centering
    \includegraphics[width=0.5\linewidth]{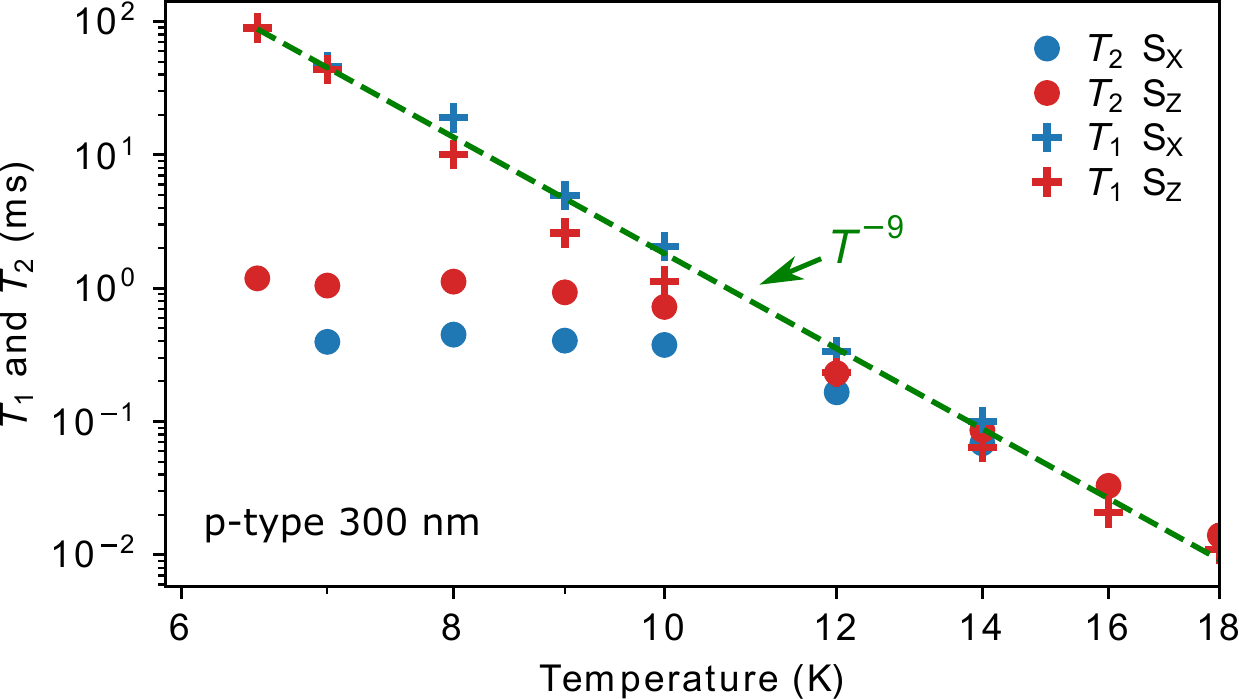}
    \caption{The temperature dependence of \tone~and \ttwo~of the high field \sx~transition and the low-field (near clock transition) \sz~transition of the deep-implanted p-type (300~nm) sample. The \tone~follows a $T^{-9}$ dependence commonly seen for silicon donors.}
    \label{fig:T1T2temp}
\end{figure}

The temperature dependence of \tone~and \ttwo~of both \sz~and \sx~transitions is presented in Fig. \ref{fig:T1T2temp}. The \tone~follows a $T^{-9}$ dependence implying that it is phonon limited as has been widely seen for donors in silicon, including selenium, as discussed in the main text. \tone~of both \sx~and \sz~transitions follows the same temperature dependence and within errors has the same values. At high temperatures, \ttwo~is limited by \tone, but as the temperature falls, \ttwo~saturates. We ensure that all measurements of \ttwo~are performed at temperatures below the saturation temperature so that no \tone~effects become mixed into the signals.

As indicated in the main text, the discrepancy between CCE simulated \ttwo~limits and experimental data suggests an additional decoherence process. We consider different noise mechanisms which could couple to our samples, but ultimately we think are unlikely to cause the extra decoherence we see in these samples. 
Three possible sources of additional noise we could find in our samples are magnetic noise from the electron spin of uncompensated boron acceptors, magnetic or electric noise from spins or traps, respectively, on the surface of the silicon, and unhealed damage from ion implantation. The first two noise sources will differently affect the samples we use in this work, and we can study the literature to estimate the effect of implantation damage.
We found similar \ttwo~times for both the deep-implanted sample (boron present, weakly affected by distant surface) and the shallow-implanted sample (no boron, close to surface), which suggests that neither the surface nor the boron are the cause of this additional noise, although we cannot rule out boron acceptors and surface proximity adding the same levels of noise. Recent measurements of Bi:$^{28}$Si with peak Bi density $\sim$50~nm below the silicon surface showed that charge noise from the surface limited coherence to $\sim$300~ms and found an effective magnetic field noise of $\sim10$~nT from the silicon surface, which would limit coherence times to $\sim$3.5~ms even at the highest values of \dfdb~\cite{ranjan2021spatially}, both much longer than timescales we measure here. 
Ion implantation has been widely used to implant $^{209}$Bi, which is approximately 2$\times$ heavier than tellurium and which requires much higher implantation energies than used here. Despite this, long coherence is routine in Bi:Si~\cite{ranjan2020multimode}, and studies on the other donors indicate that anneals for the same times as used here at 900$^{\circ}$C are already sufficient to heal implantation damage, which also suggests that this is unlikely source of the additional decoherence mechanism. 

We compare our results with previous measurements of dilute ensembles of the deep donor Se$^+$ in isotopically purified silicon, where coherence times of $\sim$80~ms were observed even away from the clock transition~\cite{nardo2015spin}. 
The predominant difference between these samples and ours are the spin density (both Se$^+$ and compensating B) and the isotopic purification. A high density of resonant spins causes instantaneous diffusion (ID) $\propto (\partial f/\partial B_0)^2$~\cite{schweiger2001pulsedEPR} and direct flip flops (dFFs)~\cite{tyryshkin2012electron}. ID should be suppressed at clock transitions and so does not explain rather short coherence we measured in this work. In addition, we performed an ID suppression experiment~\cite{tyryshkin2012electron} for the \sx~transition at X-band of the deep-implanted sample and obtained a minor increase of \ttwo~from 0.44(3) to 0.52(3)~ms, when the flip angle of the second pulse in the Hahn echo sequence changes from $\pi$ to 0. This implies that the ID limit to coherence is at much longer times than we measured here, e.g. at \dfdb~$=0.5$, we can expect ID to limit \ttwo~to about 8~ms. The active spin concentrations in our samples ($\sim 10^{16}$~cm$^{-3}$) implies, by inference from bismuth coherence times measured at clock transitions~\cite{wolfowicz2013atomic}, that dFFs would also limit coherence to $\sim 8$~ms, much longer than the coherence time we observed. These considerations suggest that our \ttwo~measurements should not be limited by spin concentrations, particularly at the \sz~clock transition.  

\begin{figure}
    \centering
    \includegraphics[width=0.5\linewidth]{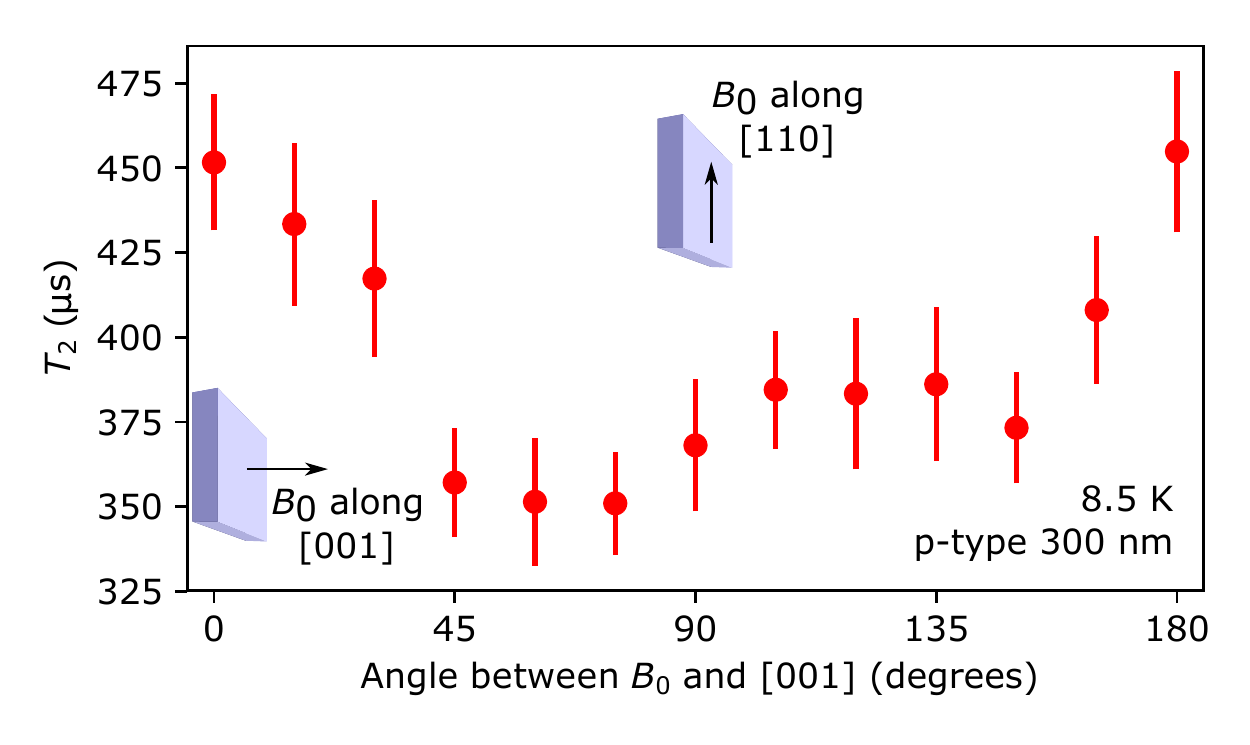}
    \caption{Angular dependence of \ttwo~of the \sx~transition of the deep-implanted p-type (300~nm) sample measured at X-band at 8.5~K.}
    \label{fig:T2_angle}
\end{figure}

The \ttwo~of donors in natural silicon also depends on the orientation of $B_0$ with respect to the principal axes of the silicon host matrix due to the anisotropy of the nuclear dipolar interaction~\cite{abe2010electron}. Maximal \ttwo\ times are observed with $B_0$ along one of the principal axes [001] (or equivalent), and a minimum is found along [111] (or equivalent). We measured the dependence of \ttwo\ of the deep-implanted sample upon the angle of $B_0$ relative to the silicon crystal axes and present the results in Fig.~\ref{fig:T2_angle}. As expected, the longest \ttwo\ is observed for $B_0$ along the [001] direction.

%Despite cubic crystal symmetry, \ttwo~is periodic in 180$^\circ$ (rather than 90$^\circ$) implying some alignment error.

\begin{figure}
    \centering
    \includegraphics[width=0.5\linewidth]{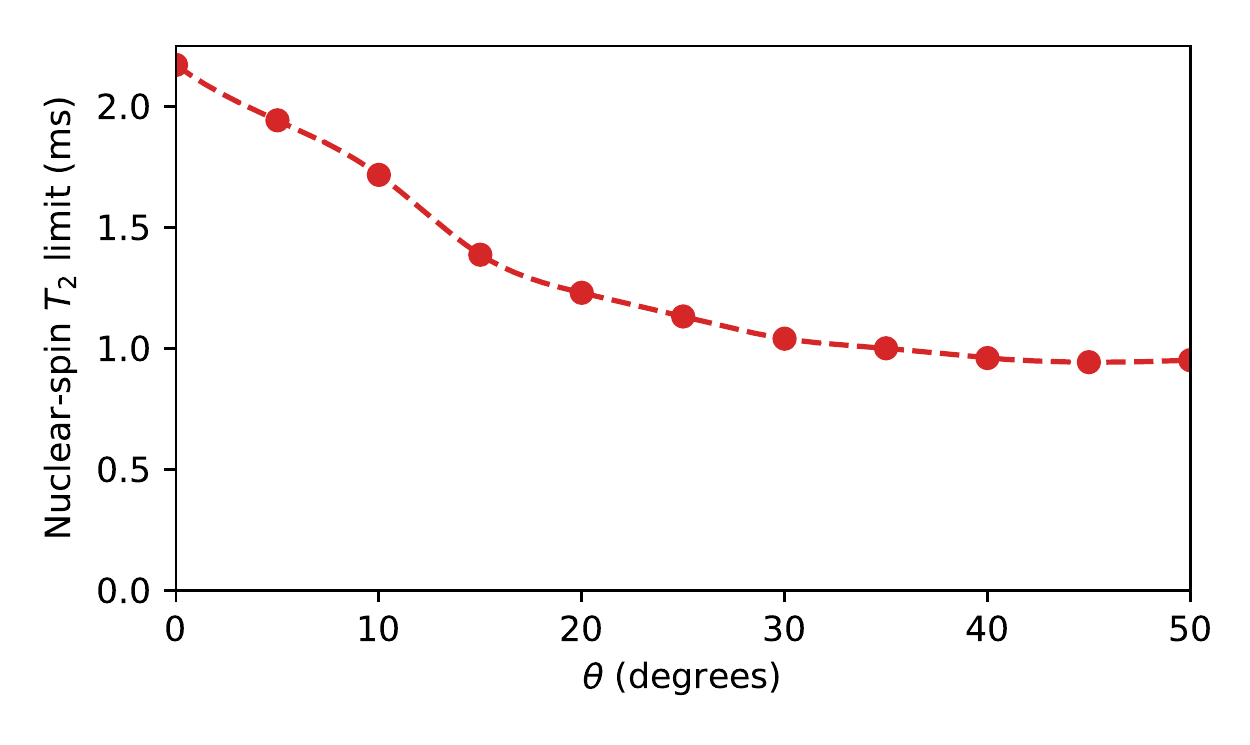}
    \caption{The angular dependence of the \ttwo~simulated by CCE-2 for the \sz~transition at 10~mT, well before the onset of ESEEM.}
    \label{fig:CCE_angle}
\end{figure}

Here we also show the CCE-2 simulated \ttwo\ times for the \sz\ transition at 10~mT revealing that the coherence time drops quite rapidly as the field is tilted away from principal crystal axis (Fig.~\ref{fig:CCE_angle}). This replicates what is seen in Fig.~\ref{fig:T2_angle}. %It implies that there is some misalignment so that, when the field is nominally along [100], there is actually a residual angle between the crystal axis and the field resulting in only partial recovery of \ttwo.

%In order to measure \sz~transitions, we had to rotate our low frequency loop-gap resonator so that all \sz~measurements are performed with an in plane field nominally along [100], whereas \sx~measurements are performed with field along [001].
%We therefore think that the reduced \ttwo~in \sz~transitions relative to the CCE-2 simulations likely arises due to misalignment of the sample to the magnetic field. We consider the \ttwo~=~1.18~ms measured for the \sz~transition at 10~mT (\dfdb$\sim0.08\gamma_{\rm e}$). 
%At this field value, the CCE-2 simulation predicts \ttwo~$\sim$2.2~ms, but, if we add in the same magnetic field noise as for \sx~transitions and a misalignment from the principal crystal axis of $\sim15^\circ$, we recover the $\sim$1.2~ms \ttwo.
%Inspecting our samples where many pieces of the same wafer are packed into a sample tube to maximise filling factor (and consequent signal) we observe a $\sim5^\circ$ misalignment. More misalignment could occur due to imperfect alignment in the home-built probe used in these experiments and so this total alignment error is plausible in our setup.

%\subsection{Prospects of $^{125}$Te:Si ESR with superconducting microresonators}
%The ESR experiments discussed in this paper translate readily to superconducting microresonators. Our use of 3D cavities here is primarily for convenience in the study of new materials with different implantation and annealing conditions, as sample exchange is straightforward in such cavities. Patterning microresonators on such implanted substrates enables the microwave magnetic field to be confined to a much smaller region in space, thus the same driving field strength can be achieved using much weaker microwave excitation. This has been shown in Bi:Si. For example, ESR studies in Refs.~[\citenum{George2010,wolfowicz2013atomic}] were performed using pulses of kW power, while Refs.~[\citenum{ranjan2020electron,ranjan2020multimode,o2020spin}] report ESR using superconducting resonators and much weaker powers.

\bibliography{bibliography}